# CL-ADMM: A Cooperative Learning Based Optimization Framework for Resource Management in MEC

Xiaoxiong Zhong, *Member*, *IEEE* , Xinghan Wang, Li Li, Yuanyuan Yang, *Fellow*, *IEEE*, Yang Qin, *senior Member*, *IEEE*, Tingting Yang, Bin Zhang, Weizhe Zhang, *senior Member*, *IEEE*

***Abstract*- We consider the problem of intelligent and efficient resource management framework in mobile edge computing (MEC), which can reduce delay and energy consumption, featuring distributed optimization and efficient congestion avoidance mechanism. In this paper, we present a Cooperative Learning framework for resource management in MEC from an Alternating Direction Method of Multipliers (ADMM) perspective, called CL-ADMM framework. First, in order to caching task efficiently in a group, a novel task popularity estimating scheme is proposed, which is based on semi-Markov process model, then a greedy task cooperative caching mechanism has been established, which can effectively reduce delay and energy consumption. Secondly, for addressing group congestion, a dynamic task migration scheme based on cooperative improved Q-learning is proposed, which can effectively reduce delay and alleviate congestion. Thirdly, for minimizing delay and energy consumption for resources allocation in a group, we formulate it as an optimization problem with a large number of variables, and then exploit a novel ADMM based scheme to address this problem, which can reduce the complexity of problem with a new set of auxiliary variables, these sub-problems are all convex problems, and can be solved by using a primal-dual approach, guaranteeing its convergences. Then we prove that the convergence by using Lyapunov theory. Numerical results demonstrate the effectiveness of the CL-ADMM and it can effectively reduce delay and energy consumption for MEC.**

This work was supported by the National Natural Science Foundation of China (Grant Nos. 61802221, 61802220, 61671165), and the Natural Science Foundation of Guangxi Province under grant 2017GXNSFAA198192, the Innovation Project of Guangxi Graduate Education under grant YCSW2019141, and the Key Research and Development Program for Guangdong Province 2019B010136001, the Peng Cheng Laboratory Project of Guangdong Province PCL2018KP005 and PCL2018KP004.
(*Corresponding authors*: Xinghan Wang, Tingting Yang).

Xiaoxiong Zhong, Xinghan Wang, Bin Zhang and Weizhe Zhang are with Peng Cheng Laboratory, Shenzhen 518000, P. R. China; Xiaoxiong Zhong is also with the Graduate School at Shenzhen, Tsinghua University, Shenzhen 518055, P. R. China. (email: {xixzhong, csxhwang}@gmail.com, {weizhe.zhang, bin.zhang}@pcl.ac.cn).

Li Li is with the Graduate School at Shenzhen, Tsinghua University, Shenzhen 518055, P. R. China. (e-mail:lilihitcs@gmail.com).

Yuanyuan Yang is with Department of Electrical and Computer Engineering, Stony Brook University, Stony Brook, NY11794, USA. (e-mail: yuanyuan.yang@stonybrook.edu).

Yang Qin is with the Department of Computer Science and Technology, Harbin Institute of Technology (Shenzhen), Shenzhen, 518055, China. (e-mail: csyqin@hit.edu.cn).

Tingting Yang is with the School of Electrical Engineering and Intelligentization, Dongguan University of Technology, Dongguan, 523000, P. R. China. (email: yangtingting820523@163.com).

## I. INTRODUCTION

With the rapid development of the wireless technology and Internet, more and more mobile terminals (MT) have different wireless access requirements for bandwidth and computing, which promotes mobile applications, such as online playing game, virtual reality, intelligent data processing and other new services continue to emerge [1-3]. However, these new mobile applications have high energy consumption and high latency, which pose a huge challenge for computing and battery capacity of the MT. Mobile Edge Computing (MEC) is a new promotion technology that supports cloud computing capabilities and edge service environments at the edge of cellular networks. The legacy base station (BS) is updated to a MEC-enabled base station (MEC-BS) by being equipped with a computing function (such as a MEC server), hence the MEC-BSs can implement MT capability enhancement, which can reduce MT application execution time and MT energy consumption. The development of these new applications and services is limited by the computing capability and battery of these MTs. If data consumption and computationally intensive tasks need to be offloaded to the cloud for execution, it can address the problem that MTs' have a lack of computing capability. However, a relatively large delay will be induced when an MT connects to the cloud over a wireless network, which is not suitable for delay-sensitive tasks.

MEC allows MT to perform computational offloading to offload its computational tasks to the MEC-BS that overwrites it. When the execution of the task at the MEC-BS is completed, the MEC-BS returns the result of the task to the MT. Due to limited computing resources, MEC-BS is unable to provide unlimited computational offload services for all tasks in the MT within coverage. Therefore, how to effectively manage MEC-BS resources (e.g., cache, energy and computational resource), maximizing system performance, is critical. On the other hand, if there are too many tasks to be uninstalled from the MT, the MEC-BS will still be overloaded. Hence, how to design an efficient resource management scheme, which can reduce delay and energy consumption, is a challenging issue.

Distributed resource allocation and caching can benefit from being deployed at the edge of network, reducing delay and energy consumption. However, cooperation and distribution for performance optimization brings to light several challenges. How could we leverage congestion for task scheduling? How can we guarantee hit ratio for a distributed caching mechanism with delay constraint? How could we guarantee convergence to a resource allocation scheme and performance optimality given task management problem?

To answer these questions, we present a Cooperative Learning framework for resource management in MEC from an Alternating Direction Method of Multipliers (ADMM) perspective, called CL-ADMM framework. CL-ADMM can leverage congestion in resource allocation among different groups (form by macro base stations, MBS), featuring distributed optimization and efficient caching mechanism, to reach intelligent and efficient resource management in MEC. The contributions of this article are as follows:

**Design contributions**. We present an intelligent resource management framework for MEC, and use cooperative learning to model cache and congestion with some constrains. Finding a distributed optimal solution for resource allocation based on ADMM with a new set of auxiliary variables, guaranteeing its convergences.

**Algorithmic contributions**. For task caching efficiently in a group, a greedy task cooperative caching algorithm based on semi-Markov process. In order to address group congestion, a dynamic task migration algorithm based on cooperative improved Q-learning is proposed, which can effectively reduce the delay and alleviate the congestion. Thirdly, for minimizing delay and energy consumption for resources allocation in a group, we present a novel ADMM based task allocation algorithm and prove its optimal value and we prove that the convergence by using the theory of Lyapunov.

**Evaluation contributions**. We evaluate both effectiveness and convergence properties of CL-ADMM from cache, group congestion, optimization in a group and their combination schemes in terms of delay, hit ratio and energy.

## II. RELATED WORK

**Task/service migration in MEC**. Recent works [4-7] propose migration schemes for MEC. [4] presented an integer programming approach for virtualized services migration with time constrain and used an efficient iterative algorithm for the optimal problem, which jointly considers migration time and computation complexity. Wang *et al.* [5] proposed a dynamic service migration in MEC based on Markov decision process, considering the distance between the user and service locations. Chen *et al.* [6] focused on the edge cognitive computing framework, which includes a key part: service migration. The services are migrated based on the behavioral cognition of a mobile user, considering migration cost and QoE using Q-learning algorithm. Zhang *et al.* [7] exploited deep reinforcement learning for task migration for MEC without users' mobility pattern in advance. Considering migration cost that characterizes the overhead incurred by migrating some workload from one edge cloud to another which includes the bandwidth cost on the network and the migration delay, Wang *et al.* [8] designed an online algorithm based on the regularization technique, which decouples the original problem into a series of subproblems that are solvable in each independent time slot, only using the solution obtained for the previous time slot as input. While above solutions do not consider congestion phenomenon and load balance in task scheduling among groups.

**Distributed resource management**. Another related set of schemes concerns the management of the shared resources [9-13]. Chen *et al.* [9] developed an optimization framework for software defined ultra-dense networks, minimizing the delay and saving the energy, which includes two sub-optimization problems: task placement and resource allocation. Consider cooperation behaviors, Xu *et al.* [10] proposed an online algorithm for service caching and task offloading in MEC systems, which jointly considers service heterogeneity, system dynamics, spatial demand and distributed coordination. Similarly, for heterogeneous services, Tan *et al.* [11] formulated a virtual resource allocation problem, and proposed ADMM algorithm to solve it, with jointly considering user association, power control and resources allocation. Zhou *et al.* [12] formulated the virtual resource allocation strategy as a joint optimization problem and used ADMM to address it, considering virtualization, caching and computing. Considering edge nodes and mobile users with time-dependent requests, Zheng *et al.* [13] proposed a convergent and scalable Stackelberg game for edge caching, and used a Stackelberg game based ADMM to solve storage allocation game and user allocation game in a distributed manner. Wang *et al.* [14] presented a parallel optimization framework which leverages the vertical cooperation among devices, edge nodes and cloud servers, and the horizontal cooperation between edge nodes by ADMM method to address the resource allocation optimal problem. Zhou *et al.* [15] studied the energy-efficient workload offloading problem and propose a low-complexity distributed solution based on consensus alternating direction method of multipliers (ADMM) for vehicular networks edge computing service provisioning. In [16], Dai *et al.* proposed a novel two-tier computation offloading framework in heterogeneous networks, which formulated joint computation offloading and user association problem for multi-task MEC system to minimize overall energy consumption through computation resource allocation and transmission power allocation. Wang *et al.* [17] proposed a heterogeneous multi-layer MEC framework, where data that cannot be timely processed at the edge are allowed to be offloaded to the upper layer MEC servers and the cloud center. Their goal is minimizing the system latency, i.e., the total computing and transmission time on all layers for the data generated by the edge devices, by jointly coordinating the task assignment, computing, and transmission resources in the framework. Li *et al.* [18] studied collaborative cache allocation and task scheduling in edge computing and modeled the task scheduling problem as a weighted bipartite graph, which can reduce latency. In [19], Alameddine *et al.* mathematically formulated the dynamic task offloading and scheduling as a mixed integer program and exploited a logic based bender decomposition approach to efficiently address the problem to optimality for low latency IoT service in MEC. For the ultra-reliable low-latency requirements in mission-critical applications, Liu *et al.* [20] proposed a new system design, where probabilistic and statistical constraints are imposed on task queue lengths, by applying extreme value theory and marrying tools from Lyapunov optimization and matching theory for user-server association, and dynamic task offloading and resource allocation. Meng *et al*. [21] studied the online deadline-aware task dispatching and scheduling in edge computing, which jointly considers with bandwidth constraint using joint optimization of networking and computing resource to meet the deadlines and proposed an online algorithm Dedas, which greedily schedules newly arriving tasks and considers whether to replace some existing tasks in order to make the new deadlines satisfied. In [22], Kim *et al*. proposed dual-side control algorithms for cost-delay tradeoff in MEC, including two optimization problems where the objective is to minimize costs subject to queue stability under two scenarios: a competition scenario and a cooperation scenario.

Recently, some learning based resource management schemes have been proposed from a artificial intelligence perspective. Ouyang *et al*. [23] proposed a novel adaptive user-managed service placement mechanism, which uses a contextual multi-armed bandit and a Thompson-sampling based online learning algorithm. However, these solutions may not be suitable in a scenario where tasks are in congestion and task collaborative behavior with arbitrary and dynamic features. In this scenario, how to design resource management framework that provides distributed and cooperative learning module and guarantees QoS requirements is a challenging issue. In [24], Tan *et al*. formulated an optimization problem for the resource allocation of communication, caching and computing in vehicular networks, and proposed the deep reinforcement learning approach with the multi-timescale framework to solve this problem which exploits mobility-aware reward estimation for the large timescale model. In [25], Sun *et al*. have studied dynamic task offloading problem in vehicular edge computing systems, and proposed an adaptive learning-based task offloading algorithm to minimize the average offloading delay, which enables each task vehicle to learn the delay performance of service vehicles in a distributed manner, without frequent exchange of state information, modifying the existing multi-armed bandit algorithms to be input-aware and occurrence-aware. Chien *et al.* [26] proposed a collaborative cache mechanism in multiple remote radio heads to multiple baseband units. They exploited Q-learning to design the cache mechanism and proposed an action selection strategy for the cache problem, finding the appropriate cache state through the reinforcement learning technology, which can reduce the traffic load of backhaul and transmission latency from the remote cloud. Different from their works, we jointly use cooperative learning to model cache and a distributed optimal solution for resource allocation based on ADMM, which is an intelligent resource management framework for large scale MEC system.

## III. CL-ADMM SYSTEM MODEL

In this section, we will describe the proposed cooperative learning system model for resource management in MEC, CL-ADMM framework, which includes three parts: *greedy task cooperative caching* based on semi-Markov process model, *dynamic task migration* based on cooperative Q-learning among groups, and *task allocation* based on ADMM in a group, as shown in Fig. 1.

**Task cooperative caching module**: caching tasks in BSs can provide low delay and improve quality of experience (QoE). Once the task has been cached in BSs, the task does not need to be uploaded and shrinks the total transmission cost. The BSs can share cached task with each other by using communications between BSs. A mobile device generates a task in which the destination BS does not cache; the destination BS will request the task from the its nearest BS. It is preferable that more task demands can be satisfied within the cooperative task cache which is close to itself. Once the mobile device request can be responded directly by the BS, the transmission cost can be minimized. Hence, a greedy task cooperative caching mechanism is proposed in GL-ADMM.

**Dynamic task migration module**: when many tasks have not cached, these tasks were uploaded to the BSs that close to itself. It may make some groups be in a congensted state with high energy consumption and delay. When a congestion occurs, the MBS can also be able to communicate with other MBSs, which are in the other group, as shown in Fig.1, especially congestion occur in a group. To address it, dynamic task migration module based on cooperative improved Q-learning is proposed, which can effectively reduce delay and alleviate congestion in a group.

**ADMM based task allocation module**: for minimizing delay and energy consumption in resources allocation, we formulate it as an optimization problem with a large number of variables. To address this problem, we propose an ADMM based scheme, reducing the complexity of problem with a new set of auxiliary variables.

The system is composed of many groups, and we denote the set of groups as $G=\{g_1, g_2, ..., g_{n-1}, g_n, g_{n+1}, ..., g_{|G|}\}$, each group is made up of a bunch of base station (BS), denoted as $BS=\{b_1, b_2, ..., b_{i-1}, b_i, b_{i+1}, ..., b_{|BS|}\}$, where $b_1$ is macro cell base station, (MBS), and $b_2$, ..., $b_i$ are small cell base stations (SBSs). Each BS is equipped with an MEC server. Each task can be separated into several separable subtasks. In this paper, we consider resource management by group mechanism, in which a group consists of an MBS, multiple SBSs, and multiple tasks. We assume the computation resource of each MEC server deployed on SBS is less than that of MBS. Each MBS have an impact on the SBSs which are in the same group, such as collecting data. Thus, if a task is offloaded to the SBS, the SBS will send the message to the MBS, the MBS can consider the next steps according to current environment. These steps include how to mitigate the congestion, how to allocate the task among the SBSs which are in the same group. The MBS can also be able to communicate with other MBSs which are in the other groups, especially congestion occurs in a group, as shown in Fig.1. Therefore, the offloading in this scenario has three steps. Firstly, the SBS or the MBS consider whether the task has been cached in it. If the task has been cached, the task needn't to be uploaded. The result will directly return to local terminal. Otherwise, the local terminal offloads its task to an MBS or a SBS which is relatively close to local terminal. The MBS will receive the message from itself or SBSs and it will determine whether this task would be processed in this group. In this step, we use the *dynamic task migration* based on cooperative Q-leaning for addressing congestion. Lastly, the offloading partition should be further divided, and MBS will determine the remaining part of task to the MEC server on the MBS or on the SBSs which has relatively higher amount of computation resources.

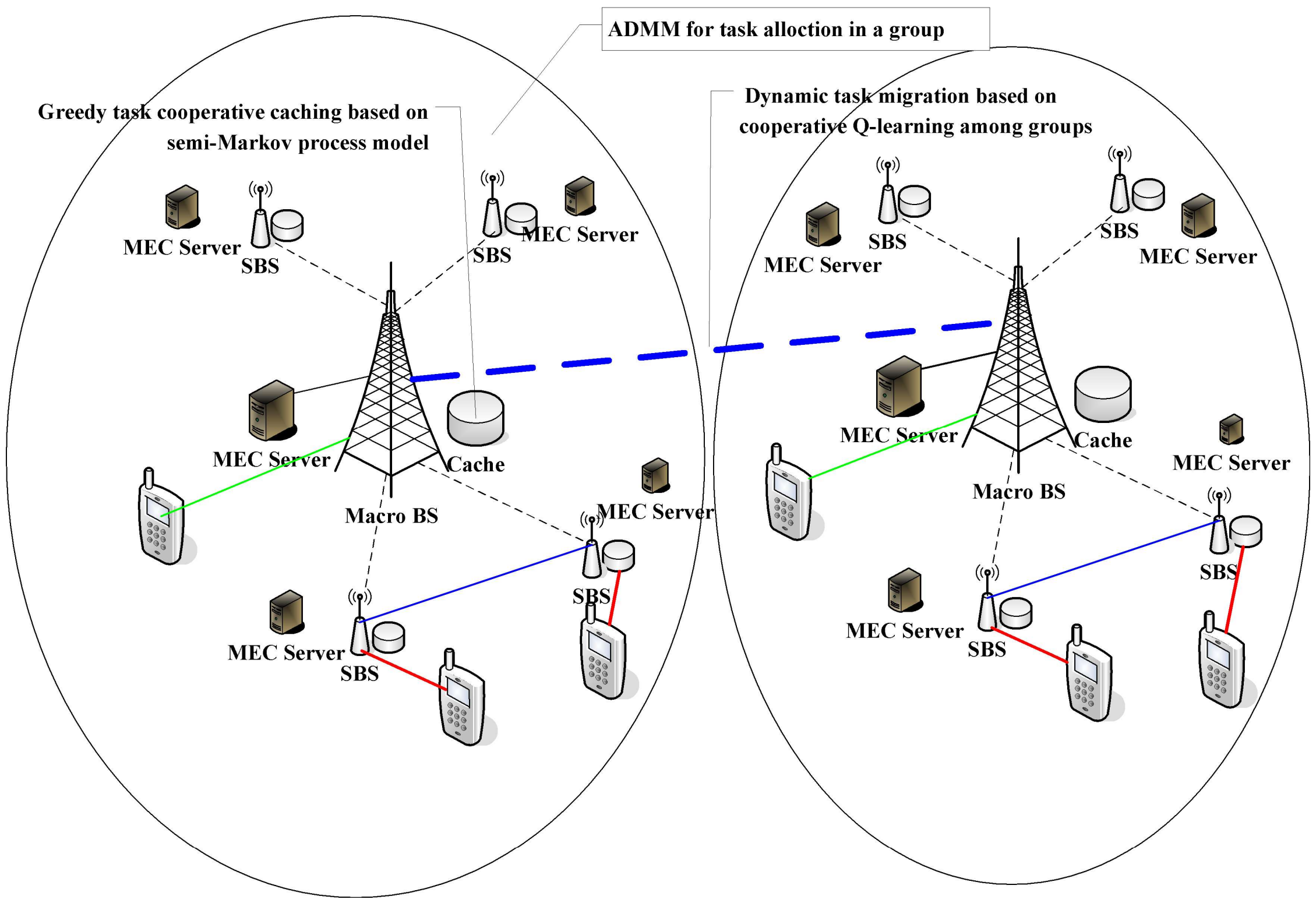

Fig. 1. CL-ADMM framework for MEC.

The set of tasks is $H = \{h_1, h_2, ..., h_{j-1}, h_j, h_{j+1}, ..., h_{|H|}\}$. Each task $h_j$ can be expressed through a vector, $h_j = \{u_j, c_j, r_j, t_j^{\max}\}$, where $u_j$ is data size of task, $c_j$ is the amount of computing resource required for the task, $r_j$ is the data size of backhauls to the local terminal, and $t_j^{\max}$ is the deadline of $h_j$. Let $M_i^c$ be the maximal value of $b_i$'s computing resource and $M_i^u$ be the maximal value of $b_i$'s storage size.

### 3.1 Computation model

In CL-ADMM framework, we describe computation model from a delay perspective. The overall delay in BS $b_i$ can be expressed as:

$$T^{(b_i,h_j)} = T_{queue}^{(b_i,h_j)} + T_{execute}^{(b_i,h_j)} + T_{down}^{(b_i,h_j)} \tag{1}$$

where $T_{queue}^{(b_i,h_j)}$ is the queuing delay, $T_{execute}^{(b_i,h_j)}$ is the calculation time and $T_{down}^{(b_i,h_j)}$ is the downing time.

$T_{queue}^{(b_i,h_j)}$ can be defined as:

$$T_{queue}^{(b_i,h_j)} = T_{wait_up}^{(b_i,h_j)} + T_{wait_execute}^{(b_i,h_j)} = \sum_{j'=1}^{j-1} T_{execute}^{(b_i,h_{j'})} + \sum_{j'=1}^{j-1} \frac{u_{j'}}{v(h_{j'})} \tag{2}$$

where the first item is the number of tasks waiting to be executed in the BS $b_i$, the second item is the number of tasks ahead of the currently task that need to be uploaded to the group, $T_{wait_up}^{(b_i,h_j)}$ is the sum of the cumulative upload time, and $T_{wait_execute}^{(b_i,h_j)}$ is the execution completion time of the previous task.

The calculation time in BS $b_i$ $T_{execute}^{(b_k,h_j)}$ can be expressed as:

$$T_{execute}^{(b_i,h_j)} = \frac{c_j}{f^{(h_j)}} \tag{3}$$

where $f^{(h_j)}$ is the maximum number of task that BS $b_i$ can handle.

The time required for task downloading $T_{down}^{(b_i,h_j)}$ can be expressed as:

$$T_{down}^{(b_i,h_j)} = \frac{r_j}{v(h_j)} \quad (4)$$

where $v(h_j)$ is that the speed of transfer between the local terminal and $b_i$.

### 3.2 Energy consumption model

In CL-ADMM framework, we exploit the following energy consumption model [27]:

$$E^{(b_i,h_j)} = E_{upload}^{(b_i,h_j)} + E_{compute}^{(b_i,h_j)} = P_U \frac{u_j}{v(h_j)} + \kappa e^{(h_j)} (f^{(h_j)})^3 T_{execute}^{(b_i,h_j)} \quad (5)$$

where $P_U$ is the transmit power, and $\kappa e^{(h_j)}$ is the computation energy efficiency coefficient of $b_i$.

### 3.3 Task priority setting

There are many tasks at a MEC-BS, however, we hope that the time difference between local execution and MEC-BS execution to be as large as possible. On the other hand, the shorter transmission time the task has, the higher priority the task has. So we jointly consider these two factors in setting the priority for the $h_j$, which can be expressed as:

$$priortiy_h_j = (1-\alpha_1)(\frac{c_j}{f_{local}} - \frac{c_j}{f^{(h_j)}}) + \alpha_1 (1/\frac{u_j}{v(h_j)}) \quad (6)$$

where $f_{local}$ is the process capability of local devices, $\alpha_1$ is a weight factor.

Based on above models, we will describe CL-ADMM framework in detail, including three components: task cooperative caching module, dynamic task migration module and ADMM based task allocation module.

## IV. GREEDY TASK COOPERATIVE CACHING BASED ON SEMI-MARKOV PROCESS

In CL-ADMM framework, there are large amount of tasks in MEC system. Hence, the state space of task is very large.[+] There are two reasons why it might be considered beneficial to use the semi-Markov method for cooperative caching. Firstly, the transition probabilities between states have the Markov property, meaning that the probability of static popularity (retention rate) transiting from current state to the highest static popularity (highest retention rate) which is independent of previous state. Another reason is that it can reduce the complexity of predictions.

In this part, we propose a cooperative caching task scheme based on semi-Markov process in MEC. As we know that it can predict the popularity of the task $h_j$ according to the historical uploading data, which can help choose a task $h_j$ based on task static popularity $s(h_j, t-\Delta t, t)$ and task retention rate $c(h_j, t-2\Delta t, t)$, selecting the task $h_j$ with high static popularity and high retention rate for caching.

$$s(h_j, t-\Delta t, t) = \frac{uploading^{(b_i,h_j)}(h_j, t-\Delta t, t)}{tol_{b_i}(t-\Delta t, t)}, \quad b_i \in BS \quad (7)$$

$$c(h_j, t-2\Delta t, t) = \frac{uploading^{(b_i,h_j)}(h_j, t-\Delta t, t)}{uploading^{(b_i,h_j)}(h_j, t-2\Delta t, t-\Delta t)}, \quad b_i \in BS \quad (8)$$

where $uploading^{(b_i,h_j)}(h_j, t-\Delta t, t)$ is the number of uploading times for task $h_j$ during time interval $(t-\Delta t, t)$ in $b_i$, $tol_{b_i}(t-\Delta t, t)$ is the number of uploading times for the all the tasks during time interval $(t-\Delta t, t)$ in $b_i$, and $uploading^{(b_i,h_j)}(h_j, t-2\Delta t, t-\Delta t)$ is the number of uploading times for task $h_j$ during time interval $(t-2\Delta t, t-\Delta t)$ in $b_i$.

We assume that the task popularity varies among tasks following the Zipf popularity model with a parameter $\xi$. We classify the popularity $s(h_j, t-\Delta t, t)$ according to a certain threshold, and each class can describe as popularity of point (POP), then we can predict the arrival of the POP. So we can predict some POP for task $h_j$, which can reduce the workload and some useless predictions. Similarly, we classify the retention rate for task $h_j$ according to a certain threshold. Thus, we can predict some retention of point (ROP) for task and then each class acts as ROP.

Next, we use POP and ROP to describe the state of the task $h_j$ for semi-Markov process model. Let $L = \{l_1, l_2, ..., l_L\}$ denote the sets of POP, arranged in descending order, $l_1$ denotes the highest static popularity. $l_L$ denotes the lowest static popularity. Let $W = \{w_1, w_2, ..., w_W\}$ denote the set of ROP, arranged in descending order, $w_1$ denotes the highest retention rate and $w_W$ denotes the lowest retention rate.

Accroding to semi-Markov process model, we have following formulars.

Let $v_{h_j}(p,q,\Delta t)$ be the first transition happens at time $x$ in state $r$.

$$v_{h_j}(p,q,\Delta t) = \begin{cases} Z_{h_j}(p,q,\Delta t) & \text{for } \Delta t = 1 \\ Z_{h_j}(p,q,\Delta t) - Z_{h_j}(p,q,\Delta t - 1) & \text{for } \Delta t > 1 \end{cases} \quad (9)$$

where $Z_{h_j}(p,q,\Delta t)$ is the task $h_j$ will transit from state $p$ to state $q$ before time unit $\Delta t$.

Hence, we have

[+] Note that we define the task as a vector, one of element is data size, and data size can be cached in BS. Hence, the tasks can be considered as caching objectives. In this paper, we design cooperative caching mechanism from a task perspective.

$$Q_{h_j}(l_{cur},l_1,\Delta t)=\begin{cases}\sum_{r=1}^{L}\sum_{x=1}^{\Delta t} v_{h_j}(l_{cur},r,x)Q_{h_j}(r,l_1,\Delta t-x) & \text{if } l_{cur}\neq l_1\\ 1-\sum_{r=1,r\neq l_{cur}}^{L} Z_{h_j}(l_{cur},r,\Delta t)+\sum_{r=1,r\neq l_{cur}}^{L}\sum_{x=1}^{\Delta t}(v_{h_j}(l_{cur},r,x))Q_{h_j}(r,l_1,\Delta t-x) & \text{if } l_{cur}=l_1\end{cases} \tag{10}$$

$$Q_{h_j}(w_{cur},w_1,\Delta t)=\begin{cases}\sum_{r=1}^{W}\sum_{x=1}^{\Delta t} v_{h_j}(w_{cur},r,x)Q_{h_j}(r,l_1,\Delta t-x) & \text{if } w_{cur}\neq w_1\\ 1-\sum_{r=1,r\neq w_{cur}}^{W} Z_{h_j}(w_{cur},r,\Delta t)+\sum_{r=1,r\neq w_{cur}}^{W}\sum_{x=1}^{\Delta t}(v_{h_j}(w_{cur},r,x))Q_{h_j}(r,l_1,\Delta t-x) & \text{if } w_{cur}=w_1\end{cases} \tag{11}$$

where $Q_{h_j}(l_{cur},l_1,\Delta t)$ is the probability that content transit from state $l_{cur}$ to $l_1$ at time $\Delta t$ , $Q_{h_j}(w_{cur},w_1,\Delta t)$ the probability that task transit from state $w_{cur}$ to $w_1$ at time $\Delta t$ and the initial condition is that:

$$Q_{h_j}(l_{cur},l_1,\Delta t)=\begin{cases}1, & l_{cur}=l_1\\ 0, & l_{cur}\neq l_1\end{cases}$$

$$Q_{h_j}(w_{cur},w_1,\Delta t)=\begin{cases}1, & w_{cur}=w_1\\ 0, & w_{cur}\neq w_1\end{cases}.$$

We update the cache placement on the task $h_j$ every interval time $\Delta t$ and use the following combined metric $p_{ij}(l_{cur},w_{cur},\Delta t)$ for task caching.

$$p_{ij}(l_{cur},w_{cur},\Delta t)=Q_{h_j}(l_{cur},l_1,\Delta t)\times Q_{h_j}(w_{cur},w_1,\Delta t),\ \ \Delta t>0 \tag{12}$$

Suppose that a caching strategy is denoted by a binary matrix $CA=\{ca_{ij}\}$ , which means that task $h_j$ cached in the local MEC server $b_i$ when $ca_{ij}=1$ ,while $ca_{ij}=0$ otherwise. Considering the local user uploads the task to the MEC server $b_i$, there exits three situations, which can be expressed through (14).

1) If the task $h_j$ had been cached in MEC server $b_i$, the task does not need to be uploaded and the uploading time will be 0, the transfer time is $(d_{ij})_1$.

2) If the task is not cached at any MEC servers, the transfer time is $(d_{ij})_3$.

3) If the task had been cached in other MEC server in a collaborative way to support service and the task had not been cached in MEC server $b_i$ (at least one MEC servers has cached task $h_j$ except for MEC server $b_i$, we choose the MEC server that is the closest $b_i$ to transfer). These MEC servers can share caching task with each other, the transfer time is $(d_{ij})_2$.

$$\begin{cases}(d_{ij})_1=0 & \textit{if task } h_j \textit{ cached in MEC server } b_i\\ (d_{ij})_3=T_{queue}^{(b_i,h_j)}\prod_{m=1}^{|BS|}(1-ca_{mj}) & \textit{if all the MEC servers in which task } h_j \textit{ is not cached}\\ (d_{ij})_2=(1-ca_{ij})(1-\prod_{m=1,m\neq i}^{|BS|}(1-ca_{mj})t_{transfer} & \textit{otherwise}\end{cases} \tag{13}$$

where

$$t_{transfer}=\frac{h_j}{v_{im}} \tag{14}$$

where $v_{im}$ is the transfer speed between $b_i$ and $b_m$ that is the closest to the $b_i$ and has task $h_j$ cached.

Let $S=\{s_1,s_2,...,s_{|BS|}\}$ denote the cache space of $BS=\{b_1,b_2,...,b_{|BS|}\}$ and $R=\{r_1,r_2,...,r_{|BS|}\}$ denote the current consumed space of $BS=\{b_1,b_2,...,b_{|BS|}\}$. Based on the aforementioned description, we can formulate a caching model based on content popularity. With objective of minimizing the average total transmission time in the proposed architecture, it can be expressed by equation (15), where $dis_{mi}$ is that the distance between $b_m$ and $b_i$

For obtaining the optimal value, we exploit a heuristic algorithm for task caching: greedy task cooperative caching based on semi-Markov process model. Let $G_{ij}$ denote the benefit of task $h_j$ has been cached in $b_i$. We can calculate the delay saving for task $h_j$ to choose caching considering a cooperative caching strategy in MEC servers. Hence, according to the situation of Equation (13), we have Equation (16) and the task caching algorithm, as shown in **Algorithm 1**.

**Algorithm 1** Greedy task cooperative caching based on semi-Markov process

**Input:** $dis_{mi}$ , $s_1,s_2,...,s_{|BS|}$ , $p_{ij}$
**Output:** matrix ***CA***
$CA=0$ , $B_j=BS$ , $r_1=0,r_2=0,...,r_{|BS|}=0$
**for** $i=1:|BS|$ **do**
  **for** $j=1:|H|$ **do**
    calculate $G_{ij}$ according the equation (16)
    **if** $G_{\max}<G_{ij}$ **then**
      $G_{\max}=G_{ij}$

```
7:      end if
8:    end for
9:  end for
10: If G_max > 0 then
11: caching task h_j to b_i , update ca_ij = 1, G_ij = 0, r_i = r_i + u_i, B_j = B_j \ b_i, CA = {ca_ij}
12: else
13:    break
14: end if
15: If r_i < s_i , then
16:   go to step 2
17: else
18:    break
19: end if
```

$$\min \sum_{j=1}^{|H|}\sum_{i=1}^{|BS|} p_{ij}\left(\left(d_{ij}\right)_1+\left(d_{ij}\right)_2+\left(d_{ij}\right)_3\right)$$
$$s.t. \tag{15}$$
$$\begin{cases} \sum_{j=1}^{|H|} ca_{ij}u_j \le s_i, & s_i \in S, b_i \in BS \\ ca_{ij} \in \{0,1\}, s_i \in S, & b_i \in BS, h_j \in H \\ m = \arg\min_{m} \left\{ dis_{mi} \middle| ca_{mj} = 1 \right\} & b_m \in BS, b_m \ne b_i \end{cases}$$

$$G_{ij} = \begin{cases} \sum_{m=1}^{|BS|} p_{mj} T_{queue}^{(b_i,h_j)} - \sum_{\substack{m=1 \\ m\ne i}}^{|BS|} p_{mj} t_{transfer}^{i->m} & if\ B_j = BS, r_i + u_j < s_i \\ \sum_{\substack{m\in B_j \\ m'\in BS-B_j}} p_{mj} t_{transfer}^{m->m'} - \sum_{\substack{m\in B_j-\{b_i\} \\ m'\in BS-B_j+\{b_i\}}} p_{jn} t_{transfer}^{m->m'} & if\ B_j \ne BS, r_i + u_i < s_i \\ 0 & otherwise \end{cases} \tag{16}$$

where $B_j$ is the set of MEC servers in which task $h_j$ is not cached. The first item in first equation indicates the profit that all the MEC servers in which task $h_j$ is not cached. The second item is that the task $h_j$ will transfer from MEC server $b_i$ to other MEC servers, when the task $h_j$ has been cached in MEC server $b_i$. The first item in second equation represents the profit that the transmission from $b_m$ to $b_{m'}$, which has cached task $h_j$ and is closest to $b_m$. The second item in second equation indicates the profit that the MEC server $b_i$ is added to the set of MEC servers in which task $h_j$ is cached.

## V. DYNAMIC TASK MIGRATION BASED ON COOPERATIVE IMPROVED Q-LEANING

In task caching module, we filter some tasks that are not satisfied given QoS requirements. However, when many tasks were uploaded to the MBSs that are closer to it, some groups may be in congestion state. In CL-ADMM framework, it uses a cooperative improved Q-learning based task migration scheme to solve it, in which modified the feedback scheme and then can effectively reduces delay and balance the resources consumption. Each group can determine its next action according to current environment in order to avoid congestion. The goal of the proposed algorithm is to balance the groups of computation capability and space capability, and alleviate the congestion in a group. When a group is congestion, the MBS will take an action under this environment. When the group is not congestion and the other one is congestion, the MBS will choose to transmit or process this task according to current environment.

In our scheme, some actions are usually composed by detecting, transmitting and processing, which are specifically defined as follows and they are setting in Table 1.

***Detecting***: When the tasks enter into the area that can be detected by the MBS or SBS, these tasks will send to the transmitting queue.

***Transmitting***: If the group is congestion, the group controller MBS will send the tasks to other group. And then it transmits them to the next group. The other group will consider process or transmit the tasks according to the environment.

***Processing***: The task can be processed in this group.

The feedback defines average reward as a relatively action which had been adopted by the group. Each MBS will obtain the positive feedback when process the tasks successfully, otherwise it will obtain the negative feedback. The value of *Processing* is setting as follows:

$$\begin{cases} +r_{proces}, & if\ it\ is\ not\ in\ congestion\ and\ process\ successfully \\ -r_{proces_1}, & if\ it\ is\ not\ in\ congestion\ and\ process\ failed \\ -r_{process_2}, & if\ it\ is\ in\ congestion\ and\ process\ successfully \\ 0, & otherwise \end{cases}$$

Table 1. Action and feedback for cooperative improved Q-learning

| Action type | Negative feedback | Positive feedback |
|---|---|---|
| *Detect.* | $-r_{detect}$ | $+r_{detect}$ |
| *Transmit.* | no | $+r_{transmit}$ |
| *Process.* | $-r_{process1}$ *or* $-r_{process2}$ | $+r_{process}$ |

We define the state of group as $GS_{state} = \left\{gs_{detect}, gs_{process}, gs_{transmit}\right\}$, $GS_{detect}$ means that the groups are probing the tasks in the sensitive area within the coverage radius. $GS_{process}$ means that the task has been processed by the groups. $GS_{transmit}$ means that the tasks has been transmitted by the groups. If the group is not in congestion, it receives a positive feedback; otherwise, it receives negative feedback, by modifying traditional cooperative Q-learning model.

In the cooperative improved Q-learning model, each group can accept the messages from the other group so as to make the best action under this situation. Hence, the Q-value function is defined as follows:

$$Q_{t+1}^{g_{n'}}(gs_t^{g_{n'}},a_t^{g_{n'}})=(1-\beta)Q_t^{g_{n'}}(gs_t^{g_{n'}},a_t^{g_{n'}})+\beta(r_{t+1}^{g_{n'}}(gs_{t+1}^{g_{n'}}) +\gamma\sum_{g^n\in G}f(g^n)V_t^{g_n}(gs_{t+1}^{g_{n'}})) \quad (17)$$

$$V_t^{g_n}(gs_{t+1}^{g_{n'}})=\max_{a\in A}Q_t^{g_n}(gs_{t+1}^{g_{n'}},a)$$

where $gs_t^{g_{n'}}\in GS_{state}$ represents the state as time $t$, $a_t^{g_{n'}}$ represents the current selected action; $r_{t+1}^{g_{n'}}(gs_{t+1}^{g_{n'}})$ is the feedback, which is received at time $t+1$ after taking action $a_t^{g_{n'}}$. $V_t^{g_n}(gs_{t+1}^{g_{n'}})$ is the state value function of the other group $g_n$ at time $t$, $f(g^n)$ is the group $g^n$ 's weight factor, and $\gamma$ is a discounted factor.

Each group learns from the other groups with the purpose of finding the optimal policy on all states, which is conducive to itself and its neighboring groups. The value function is defined as follows:

$$V_{g_{n'}}(gs_{g_{n'}})=R(gs_{g_{n'}},a)+\gamma\sum_{g_n\in G}f(g_n)P\left(gs_{g_{n'}},gs'_{g_{n'}},a\right)V_{g_n}(gs'_{g_{n'}}) \quad (18)$$

where $R(gs'_{g_{n'}},a)$ is the reward, and $P\left(gs_{g_{n'}},gs'_{g_{n'}},a\right)$ is the probability that the group $g_n$ takes action $a$, making it enter into the next state $gs'_{g_{n'}}$ from previous state $gs_{g_{n'}}$ .

The optimal policy can be expressed as:

$$\pi_{g_{n'}}(gs_{g_{n'}})=\arg\max_{a\in Action}\left[R(gs_{g_{n'}},a)+\gamma\sum_{g_n\in G}f(g_n)P\left(gs_{g_{n'}},gs'_{g_{n'}},a\right)V_{g_n}(gs'_{g_{n'}})\right] \quad (19)$$

An $|G|$ -dimensional vector $Num=(num_1,num_2...,num_{|G|})$ is used to denote the number of tasks in each group, where $num_i$ is the number of tasks in the $g^{th}$ group currently. $P(num_g)$ is the proportion of the tasks of the $g^{th}$ group in the total number of tasks .

$$P(num_g)=\frac{num_g}{\sum_{g=1}^{|G|}num_g} \quad (21)$$

In order to solve the congestion problem, it is necessary to balance the number of task in each group during the process of task migration. In addition, we should determine that whether the action is chosen in the right way. Hence, we exploit current entropy of the groups to address the problem, which can be expressed:

$$F_{cur}=-\sum_{i=1}^{|G|}P(num_g)\log P(num_g) \quad (22)$$

Hence, the dynamic task migration algorithm can be listed as **Algorithm 2.**

**Algorithm 2** Dynamic task migration based on cooperative Q-learning

**1**: **Initialize** Q($gs$, $a$), $F_{\max}$ , $\gamma$ , $t$ (the time slot), $T$ (the maximal time slot)
**2**: At time $t$, group $g_n$ determines its current state $gs_t^{g_n}$
**3**: determine the exploration probability $\varepsilon_t$ according to (21)
**4**:: **while** $t\le T$ **do**
**5**: Generate a random number $\delta(\delta\in(0,1))$
**6**: Select the optimal action $a^p$ according to (19)
**7**: Generate a random action $a^{random}$
**8**: **if** $\delta\le\varepsilon_t$ **then**
**9**: $a^{g_{n'}}=a^p$
**10: else**
**11**: $a^{g_{n'}}=a^{random}$
**12**: **end if**
**13**: calculate $F_{cur}$ according to (22)
**14**: **if** $F_{\max}<F_{cur}$ **then**
**15**: $F_{\max}=F_{cur}$
**16: end if**
**17: end while**
**18**: update reward table according to (17)

## VI. ADMM FOR RESOURCE ALLOCATION IN A GROUP

In this section, we will describe ADMM based resource optimization in a group, which jointly considers delay and energy in optimization.

First, the queuing time in equation (2) should be updated according to the following equation:

$$T_{queue}^{(b_i,h_j)}=T_{wait_up}^{(b_i,h_j)}+T_{wait_execute}^{(b_i,h_j)}=\sum_{j'=1}^{j-1}T_{execute}^{(b_{i'},h_n)}\times x_{ij'}+\sum_{j'=1}^{j}\frac{u_{j'}}{v(h_{j'})}\times x_{ij'} \quad (23)$$

where $x_{ij'}$ is denotes the probability that the task $h_{j'}$ select the base station $b_i$ in a group, and it has $\sum_{i=1}^{|BS|}x_{ij}=1$.

There are two main reasons involved why the original ADMM can't be simply applied to solve our problems. Firstly, due to the fact that problem in the original ADMM are usually decomposable, but in our problem all the variables are tightly coupled. Secondly, the original ADMM has high complexity and low scalability. To tackle these challenges, we introduce CL-ADMM framework with a novel ADMM scheme. It has $|H|\times|BS|$ variables in our optimization problem, where $|BS|$ is the number of BS, $|H|$ is the number of the tasks. Thus, the computational complexity significantly increases as the number of tasks and BSs. To solve the large-scale optimization problem, we propose a scalable and practical distributed method, by ADMM with blending the advantage of the auxiliary variable, and use the ADMM with Gaussian back substitution in order to improve the convergence of the multi-block ADMM.

Hence, we formulate the cost-minimizing problem by equation (24), in which variables are connected to above equations and are capacity constraint. To improve this situation, we formulate the optimization problem by introducing a new set of auxiliary variable $y_{ij}$, which meets the following constraints in equation (25).

Clearly, the new formulation is now separable over the two sets of variables $x$ and $y$. The augmented Lagrangian associated is defined as (26), where $\lambda_{ij}$ is the Lagrangian multiplier and $\rho$ is the penalty parameter.

With augmented Lagrangian, we can conclude that:

$$\begin{aligned}&\min L_\rho(x,y,\lambda)\\&s.t.\ \text{Equation (25-1),(25-2),(25-3),(25-4),(25-5),(25-6),(25-8)}\end{aligned}\tag{27}$$

Since the objective is a linear function and will be strictly convex. The iterative scheme of ADMM into iterations of $x, y$ can be expressed as:

$$\begin{cases}x^{k+1}=\arg\min\limits_{x} L_\rho(x,y^k,\lambda^k)\\ y^{k+1}=\arg\min\limits_{y} L_\rho(x^{k+1},y,\lambda^k)\\ \lambda^{k+1}=\lambda^k+\rho(x^{k+1}-y^{k+1})\end{cases}\tag{28}$$

According to Equation (28), we can conclude that:

$$\begin{aligned}&\min\ L_\rho(x,y^k,\lambda^k)\\&s.t.\ \text{Equation (25-1),(25-2),(25-3),(25-5)}\end{aligned}\tag{29}$$

Next, we use the ADMM with Gaussian back substitution to improve the convergence of the proposed scheme. The following sub-problem needs to be independently solved by (30).

At the $(k+1)^{\text{th}}$ iteration, the optimal solution to the sub-problem in above equation can be expressed as equation (31).

The Karush-kuhn-Tucker (KKT) can be expressed by equation (32), where $\beta_{ij}\ge 0,\gamma_{ij}\ge 0,\varepsilon\ge 0$ and $\delta_i\ge 0$ are Lagrangian multiplies.

$$\min\ utility=\sum_{i=1}^{|BS|}\sum_{j=1}^{|H|}\left(coe\left(\sum_{j'=1}^{j-1}T_{execute}^{(b_i,h_{j'})}\times x_{ij'}+\sum_{j'=1}^{j}\frac{u_{j'}}{v(h_{j'})}\times x_{ij'}+(T_{execute}^{(b_i,h_j)}+T_{down}^{(b_i,h_j)})x_{ij}\right)+(1-coe)E^{(b_i,h_j)}x_{ij}\right)$$

$$s.t\begin{cases}\sum\limits_{i=1}^{|BS|}\left(\sum\limits_{j'=1}^{j-1}T_{execute}^{(b_i,h_{j'})}\times x_{ij'}+\sum\limits_{j'=1}^{j}\frac{u_{j'}}{v(h_{j'})}\times x_{ij'}+(T_{execute}^{(b_i,h_j)}+T_{down}^{(b_i,h_j)})x_{ij}\right)\le t_j^{\max},\ 0\le coe\le 1\\ 0\le x_{ij}\le 1,\ \sum\limits_{i=1}^{|BS|}x_{ij}=1,\ \sum\limits_{j=1}^{|H|}c_jx_{ij}\le M_i^c,\ \forall M_i^c\in M_c,\ \sum\limits_{j=1}^{|H|}u_jx_{ij}\le M_i^u,\ \forall M_i^u\in M_u, j\in\{1,2,3...,|H|\}\end{cases}\tag{24}$$

$$\min\ utility=\sum_{i=1}^{|BS|}\sum_{j=1}^{|H|}\left(coe\left(\sum_{j'=1}^{j-1}T_{execute}^{(b_i,h_{j'})}\times x_{ij'}+\sum_{j'=1}^{j}\frac{u_{j'}}{v(h_{j'})}\times x_{ij'}+(T_{execute}^{(b_i,h_j)}+T_{down}^{(b_i,h_j)})x_{ij}\right)+(1-coe)E^{(b_i,h_j)}x_{ij}\right)$$

$$s.t\begin{cases}\sum\limits_{i=1}^{|BS|}\left(\sum\limits_{j'=1}^{j-1}T_{execute}^{(b_i,h_{j'})}\times x_{ij'}+\sum\limits_{j'=1}^{j}\frac{u_{j'}}{v(h_{j'})}\times x_{ij'}+(T_{execute}^{(b_i,h_j)}+T_{down}^{(b_i,h_j)})x_{ij}\right)\le t_j^{\max} & \text{(25-1)}\\ 0\le x_{ij}\le 1,\ 0\le coe\le 1 & \text{(25-2)}\\ \sum\limits_{j=1}^{|H|}c_jx_{ij}\le M_i^c\quad \forall M_i^c\in M_c & \text{(25-3)}\\ \sum\limits_{j=1}^{|H|}u_jy_{ij}\le M_i^u\quad \forall M_i^u\in M_u & \text{(25-4)}\\ \sum\limits_{i=1}^{|BS|}x_{ij}=1\quad j\in\{1,2,3...,|H|\} & \text{(25-5)}\\ 0\le y_{ij}\le 1 & \text{(25-6)}\\ x_{ij}=y_{ij} & \text{(25-7)}\\ \sum\limits_{i=1}^{|BS|}y_{ij}=1\quad j\in\{1,2,3...,|H|\} & \text{(25-8)}\end{cases}\tag{25}$$

$$L_\rho(x,y,\lambda)=\sum_{i=1}^{|BS|}\sum_{j=1}^{|H|}(coe(\sum_{j'=1}^{j-1}T_{execute}^{(b_i,h_{j'})}\times x_{ij'}+\sum_{j'=1}^{j}\frac{u_{j'}}{v(h_{j'})}\times x_{ij'}+(T_{execute}^{(b_i,h_j)}+T_{down}^{(b_i,h_j)})x_{ij})+(1-coe)E^{(b_i,h_j)}x_{ij})+\sum_{i=1}^{|BS|}\sum_{j=1}^{|H|}\lambda_{ij}(x_{ij}-y_{ij})+\frac{\rho}{2}(x_{ij}-y_{ij})^2 \tag{26}$$

$$\begin{aligned}&\min\ L_\rho(x,y^k,\lambda^k,v^k)=coe(\sum_{j'=1}^{j-1}T_{execute}^{(b_i,h_{j'})}\times x_{ij'}+\sum_{j'=1}^{j}\frac{u_{j'}}{v(h_{j'})}\times x_{ij'}+(T_{execute}^{(b_i,h_j)}+T_{down}^{(b_i,h_j)})x_{ij})+(1-coe)E^{(b_i,h_j)}x_{ij}+(|H|-j)\left\{T_{execute}^{(b_i,h_j)}x_{ij}+\frac{u_j}{v(h_j)}x_{ij}\right\}+\\&\lambda_{ij}^k(x_{ij}-y_{ij}^k)+\frac{\rho}{2}(x_{ij}-y_{ij}^k)^2+v_j^k(\sum_{m=1}^{i-1}\tilde{x}_{mj}^{k+1}+x_{ij}+\sum_{m=i+1}^{|BS|}x_{mj}^{k+1}-1)+\frac{\rho}{2}(\sum_{m=1}^{i-1}\tilde{x}_{mj}^{k+1}+x_{ij}+\sum_{m=i+1}^{|BS|}x_{mj}^{k+1}-1)^2\\&s.t.(25\text{-}1)(25\text{-}2)(25\text{-}3),i\in\{1,2,3,...,|BS|\},j\in\{1,2,3...,|H|\}\end{aligned} \tag{30}$$

$$\begin{aligned}&\psi(x,\varepsilon,\beta,\gamma,\delta)=L_\rho(x,y^k,\lambda^k,v^k)-\beta_{ij}x_{ij}+\gamma_{ij}(x_{ij}-1)+\delta_i(\sum_{j=1}^{|H|}c_jx_{ij}-M_i^c)+\\&\varepsilon(\sum_{i=1}^{|BS|}(\sum_{j'=1}^{j-1}T_{execute}^{(b_i,h_{j'})}\times x_{ij'}+\sum_{j'=1}^{j}\frac{u_{j'}}{v(h_{j'})}\times x_{ij'}+(T_{execute}^{(b_i,h_j)}+T_{down}^{(b_i,h_j)})x_{ij})\text{-}t_j^{\max})\end{aligned} \tag{31}$$

$$\begin{cases}\dfrac{\partial\psi(x,\varepsilon,\beta,\gamma,\delta)}{\partial x}=0,\ \gamma_{ij}(x_{ij}-1)=0\ ,\ \gamma_{ij}\ge 0,\\ \beta_{ij}(\text{-}x_{ij})=0,\ \beta_{ij}\ge 0,\\ \varepsilon(\sum_{i=1}^{|BS|}(\sum_{j'=1}^{j-1}T_{execute}^{(b_i,h_{j'})}\times x_{ij'}+\sum_{j'=1}^{j}\frac{u_{j'}}{v(h_{j'})}\times x_{ij'}+(T_{execute}^{(b_i,h_j)}+T_{down}^{(b_i,h_j)})x_{ij})\text{-}t_j^{\max})=0,\ \varepsilon\ge 0,\\ \delta_i(\sum_{j=1}^{|H|}c_jx_{ij}-M_i^c)=0,\ \delta_i\ge 0.\end{cases} \tag{32}$$

$$\begin{cases}\tilde{x}_{ij}=\dfrac{-(1-coe)\left\{E^{(b_i,h_j)}+(|H|\text{-}j)\left(E^{(b_i,h_j)}\right)\right\}-coe\left\{\dfrac{u_j}{v(h_j)}+T_{execute}^{(b_i,h_j)}+T_{down}^{(b_i,h_j)}+(|H|\text{-}j)\left(T_{execute}^{(b_i,h_j)}+\dfrac{u_j}{v(h_j)}\right)\right\}}{2\rho}+\\ \dfrac{-\lambda_{ij}^k+\rho y_{ij}^k-v_j^k-\rho(\tilde{x}_{1j}^{k+1}+...+\tilde{x}_{i-1j}^{k+1}+x_{i+1j}^{k+1}+...+x_{|BS|j}^{k+1}-1)+\beta_{ij}\text{-}\gamma_{ij}\text{-}\varepsilon\left(\dfrac{u_j}{v(h_j)}+T_{execute}^{(b_i,h_j)}+T_{down}^{(b_i,h_j)}\right)\text{-}\delta_ic_j}{2\rho}\\ \dot{\gamma}_{ij}=(x_{ij}-1)\\ \dot{\beta}_{ij}=\text{-}x_{ij}\\ \dot{\varepsilon}=\sum_{i=1}^{|BS|}(\sum_{j'=1}^{j-1}T_{execute}^{(b_i,h_{j'})}\times x_{ij'}+\sum_{j'=1}^{j}\frac{u_{j'}}{v(h_{j'})}\times x_{ij'}+(T_{execute}^{(b_i,h_j)}+T_{down}^{(b_i,h_j)})x_{ij})\text{-}t_j^{\max}\\ \dot{\delta}_i=(\sum_{j=1}^{|H|}c_jx_{ij}-M_i^c)\\ \dot{v}_j=\sum_{i=1}^{|BS|}x_{ij}-1\end{cases} \tag{33}$$

$$\begin{aligned}\min\ L_\rho(x^{k+1},y^k,\lambda^k,z^k) =\ & coe\Big(\sum_{j'=1}^{j-1} T_{execute}^{(b_i,h_{j'})}\times x_{ij'} + \sum_{j'=1}^{j}\frac{u_{j'}}{v(h_{j'})}\times x_{ij'} + (T_{execute}^{(b_i,h_j)} + T_{down}^{(b_i,h_j)})x_{ij}^{k+1}\Big) + (1-coe)E^{(b_i,h_j)}x_{ij}^{k+1}\Big)+\\ & (|H|-j)\left\{T_{execute}^{(b_i,h_j)}x_{ij}^{k+1} + \frac{u_j}{v(h_j)}x_{ij}^{k+1}\right\} + \lambda_{ij}^k(x_{ij}^{k+1}-y_{ij}^k) + \frac{\rho}{2}(x_{ij}^{k+1}-y_{ij}^k)^2 + z_j^k\Big(\sum_{m=1}^{i-1}\tilde{y}_{mj}^{k+1} + y_{ij} + \sum_{m=i+1}^{|BS|}y_{mj}^{k+1} - 1\Big) + \frac{\rho}{2}\Big(\sum_{m=1}^{i-1}\tilde{y}_{mj}^{k+1} + y_{ij} + \sum_{m=i+1}^{|BS|}y_{mj}^{k+1} - 1\Big)^2\\ & s.t.(25\text{-}4)(25\text{-}6), i\in\{1,2,3,...,|BS|\}, j\in\{1,2,3...,|H|\},\ 0\le coe\le 1\end{aligned} \tag{34}$$

These problems are all convex problems, which can be solved by using a primal-dual approach. The sub-problems can be solved by (33). Moreover, once $x^{k+1}$ is obtained, the problem $y^{k+1}$ can be solved in a similar way, which can be solved by (34).

The Lagrangian of (34) and the KKT can be expressed by (35) and (36).

$$\psi(y,\mu,\varsigma,\vartheta) = L_\rho(x^{k+1},y^k,\lambda^k,z^k)\text{-}\mu_{ij}y_{ij} + \varsigma_{ij}(y_{ij}-1) + \vartheta_i\Big(\sum_{j=1}^{|H|}u_jy_{ij} - M_i^u\Big) \tag{35}$$

$$\begin{cases}\dfrac{\partial\psi(y,\mu,\varsigma,\vartheta)}{\partial y} = 0,\\ \varsigma_{ij}(y_{ij}-1)=0\ ,\ \ \varsigma_{ij}\ge 0,\\ \mu_{ij}(-y_{ij})=0,\ \ \mu_{ij}\ge 0,\\ \vartheta_i\Big(\sum_{j=1}^{|H|}u_jx_{ij} - M_i^u\Big)=0,\ \ \vartheta_i\ge 0.\end{cases} \tag{36}$$

where $\mu_{ij}\ge 0, \varsigma_{ij}\ge 0$, and $\vartheta_i\ge 0$ are Lagrangian multiplies.

Similar to (33), we have equation (37). Next, we will describe ADMM with Gaussian back substitution correcting the output [28]. We first define some vectors in equation (38) and matrix $O = diag(\rho A_2^T A_2, ..., \rho A_{|BS|}^T A_{|BS|}, \frac{1}{\rho}I_m)$. We denote another matrix $M$ in the equation (42).

The proposed Gaussian back substitution correction step can be expressed in (43).

$$\begin{cases}\tilde{y}_{ij} = \dfrac{\lambda_{ij}^k + \rho x_{ij}^{k+1} - z_j^k - \rho(\tilde{y}_{1j}^{k+1} + ... + \tilde{y}_{i-1j}^{k+1} + y_{i+1j}^{k+1} + ... + y_{|BS|j}^{k+1}\text{-}1) + \mu_{ij}\text{-}\varsigma_{ij}\text{-}\vartheta_i u_j}{2\rho}\\ \dot{\varsigma}_{ij} = (y_{ij}-1)\\ \dot{\mu}_{ij} = \text{-y}_{ij}\\ \dot{\vartheta}_i = \Big(\sum_{j=1}^{|H|}u_jx_{ij} - M_i^u\Big)\\ \dot{z}_j = \sum_{i=1}^{|BS|}y_{ij} - 1\end{cases} \tag{37}$$

$$\begin{aligned}V_x &= (x_{2j}, x_{3j}, ..., x_{|BS|j}, v_j)\ \ j\in\{1,2,3...,|H|\}\\ \tilde{V}_x &= (\tilde{x}_{2j}, \tilde{x}_{3j}, ..., \tilde{x}_{|BS|j}, \tilde{v}_j)\ \ j\in\{1,2,3...,|H|\}\end{aligned} \tag{38}$$

$$M = \begin{Bmatrix}\rho A_2^T A_2 & 0 & \cdots & \cdots & 0\\ \rho A_3^T A_2 & \rho A_3^T A_3 & \ddots & & \vdots\\ \vdots & \vdots & \ddots & \ddots & \vdots\\ \rho A_{|BS|}^T A_2 & \rho A_{|BS|}^T A_3 & \cdots & \rho A_{|BS|}^T A_{|BS|} & 0\\ 0 & 0 & \cdots & 0 & \frac{1}{\rho}I_m\end{Bmatrix} \tag{39}$$

where $A_2 = A_3 = ... = A_{|BS|} = 1$.

$$O^{-1}M^T(V_x^{k+1} - V_x^k) = \alpha_2(\tilde{V}_x^k - V_x^k) \tag{40}$$

where the matrix $O^{-1}M^T$ is upper-triangular block matrix and $\alpha_2$ is a corrector.

Similarly, we define following vectors in equation (41) and have (42). Hence, we can obtain **Theorem 1**.

$$\begin{aligned}V_y &= (y_{2j}, y_{3j}, ..., y_{|BS|j}, z_j)\ \ j\in\{1,2,3...,|H|\}\\ \tilde{V}_y &= (\tilde{y}_{2j}, \tilde{y}_{3j}, ..., \tilde{y}_{|BS|j}, \tilde{z}_j)\ \ j\in\{1,2,3...,|H|\}\end{aligned} \tag{41}$$

$$O^{-1}M^T(V_y^{k+1} - V_y^k) = \alpha_2(\tilde{V}_y^k - V_y^k) \tag{42}$$

**Theorem 1** *The proposed ADMM algorithm is convergence from a hierarchical perspective.*

Proof: Please refer to **Appendix A**.

Hence, the task allocation for MEC can be described as **Algorithm 3**, which is based on modified ADMM with Gaussian back substitution.

---

**Algorithm 3** Task allocation for MEC based on ADMM in a group

---

**Input:**

The space size of MEC server $M_i^u$ , $\forall M_i^u \in M_u$

The amount of computing resource in per MEC server: $M_i^c$ , $\forall M_i^c \in M_c$ ,

Parameter setting: $\alpha_2 \in (0,1), \rho$

The rounds of iterations: $\kappa$

**Output**: $x_{ij}$

1: **Initialize** $x^0, \tilde{x}^0, y^0, \tilde{y}^0,, z_j^0, \tilde{z}_j^0, v_j^0, \tilde{v}_j^0$

2: **while** $k < \kappa$ **do**

3: **for** $j = 1,2,....,|H|$ **do**

4: **for** $i = 1,2,...,|BS|$ **do**

5: calculate $\tilde{x}_{ij}$ , $(\gamma_{ij}^{k+1})^+$, $(\beta_{ij}^{k+1})^+$ according to the given the variable $y_{ij}^k$ and $\lambda_{ij}^k$ , and (33)

6: **end for**

calculate $\left(\varepsilon^{k+1}\right)^+$, $v_j^{k+1}$ according to (33)
Gaussian back substitution correction for $x$ according to (40), and
$x_{1j}^{k+1}=\tilde{x}_{1j}^{k}$
**end for**
calculate $\left(\delta_i^{k+1}\right)^+$ according to (33)
**for** $j=1,2,....,|H|$ **do**
**for** $i=1,2,...,|BS|$ **do**
calculate $\tilde{y}_{ij}$, $\left(\varsigma_{ij}^{k+1}\right)^+$, $\left(\mu_{ij}^{k+1}\right)^+$ according to the given variable $\tilde{x}_{ij}^{k+1}$,
$\lambda_{ij}^{k}$ and (37)
**end for**
calculate $z_j^{k+1}$ according to (37)
Gaussian back substitution correction for $y$ according to (42), and
$y_{1j}^{k+1}=\tilde{y}_{1j}^{k}$
**end for**
calculate $\left(\vartheta_i^{k+1}\right)^+$ according to (37)
update $\lambda$ according to (28)
$k++$
**end while**

## VII. PERFORMANCE EVALUATION

In this part, we evaluate the performance of CL-ADMM framework. The parameters setting are shown in Table 2.

Table 2. Simulation parameters setting.

| Parameter | Value |
|---|---|
| The transmit power of user | 0.1w |
| Input data size | 5 kbits-10kbits |
| Delay threshold | 15-30s |
| Computational capability | 10-100GHz CPU cycles/s |
| The path loss exponent | 4 |
| The bandwidth | 20MHz |
| Computation workload | 18000 CPU cycles/bit |
| Computation efficiency coefficient | $10^{-26}$ |
| The power of noise | -172dBm/Hz |
| The transmit power of BS | 40 |
| BS coverage area | 200m*200m |
| Max group | 3 |
| $\alpha_1,\alpha_2$ | 0.5 |
| $M_i^u$ | 100M |
| $\beta$ | 0.1 |
| $T$ | 1000 |

As shown in Fig. 2, the number of the task is 5000. The number of BS is 50. Our solution is better than the schemes without cooperative caching in terms of hit ratio. The proposed scheme can more effectively detect the cache state in caching task and uses a cooperative learning based scheme for caching task.

In Fig.3, the number of the task is 600. The number of BS is 50. The number of task types is 10. The task cooperative caching module will filter some tasks that are not satisfied given QoS requirements. Hence, we study the impact of the no-caching task on utility. As expected, we observe that the utility increases with increasing the number of the no-caching tasks for the following schemes: cooperative caching, cooperative caching + migration and cooperative caching + migration + ADMM. However, the combination of the proposed scheme (cooperative caching + migration + ADMM) obtains a better performance. The reason is that the combination scheme jointly considers task cooperative caching, congestion avoidance and task optimization scheduling from a learning perspective, which is more effective to allocate resource.

In Fig.4, we present the convergence of the ADMM based-algorithm (the parameter is 2) with different number of tasks. The number of task types is 10. The number of BS is 3. The utility will increase with increasing the number of tasks and then enter a stable status within the first 30 iterations.

The Fig.5 shows the increase trend in hit ratio in cooperative caching between 10% of buffer size and 80% of buffer size. The number of tasks is 50000, the type of tasks is 50, and the number of BS is 10. The buffer size is 50*10kbits. The hit ratio increases dramatically from 0.200502 to 0.930233. By contrast, the hit ratio in random caching fluctuates with an overall slightly upward trend. The reason is that the hit ratio has a rise trend along with increase of buffer size.

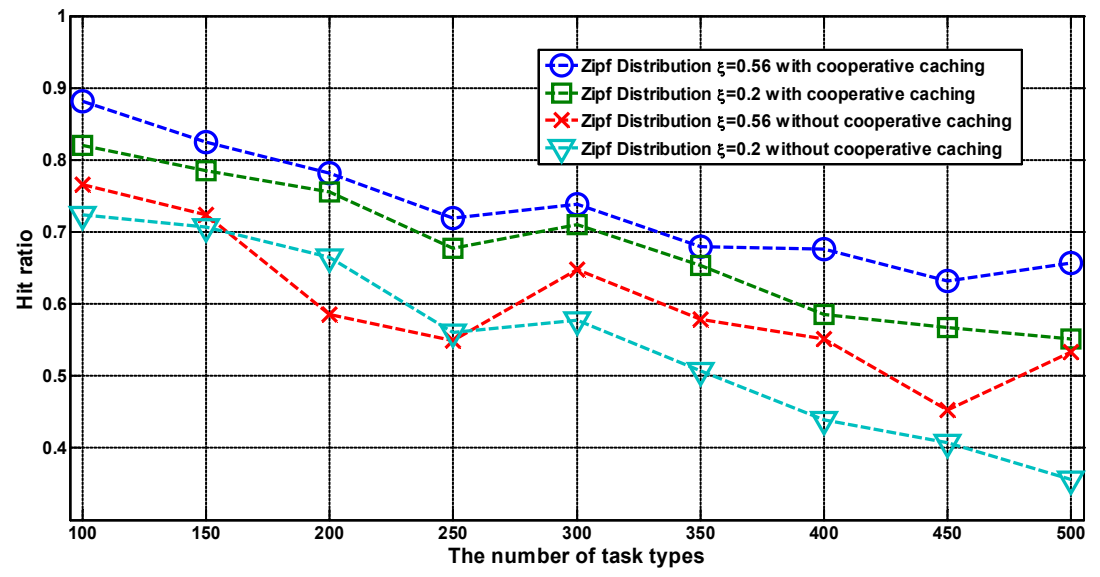

Fig. 2 The number of task types vs. hit ratio.

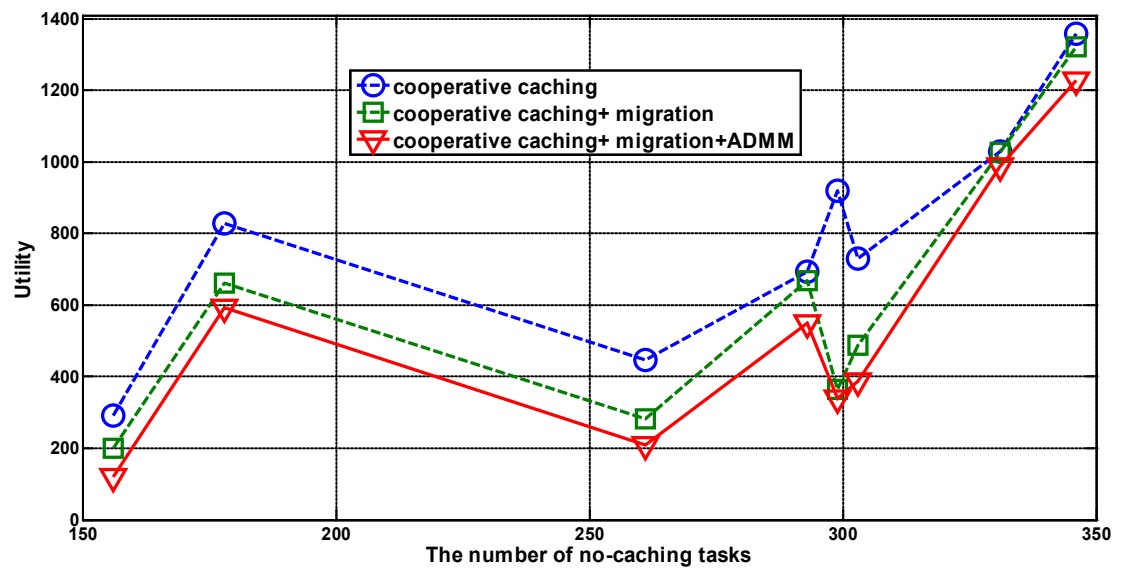

Fig. 3. The number of no-caching tasks vs. utility.

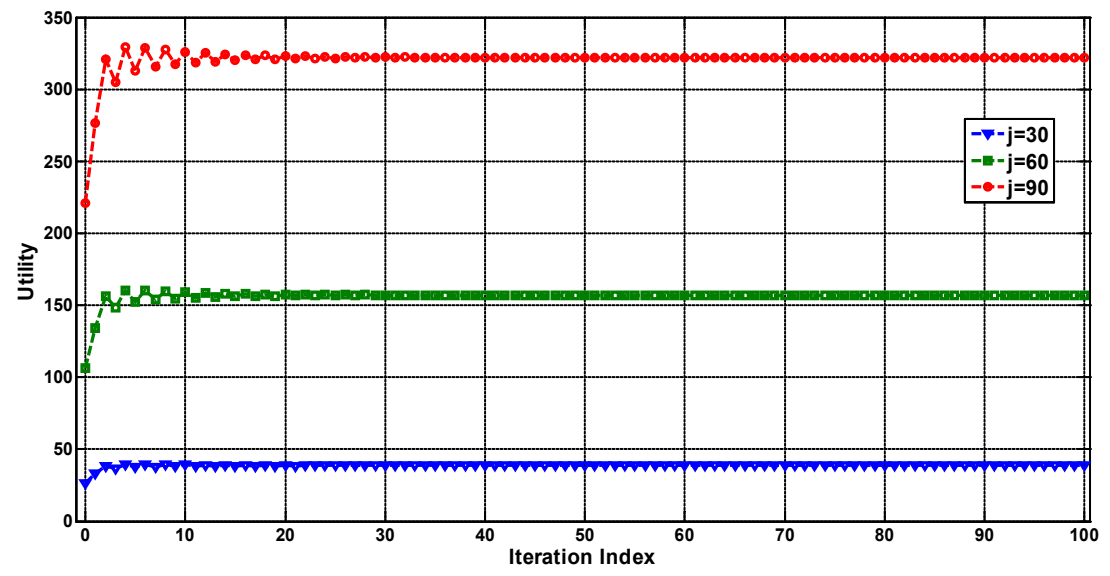

Fig. 4. Convergence progresses of ADMM-based algorithm.

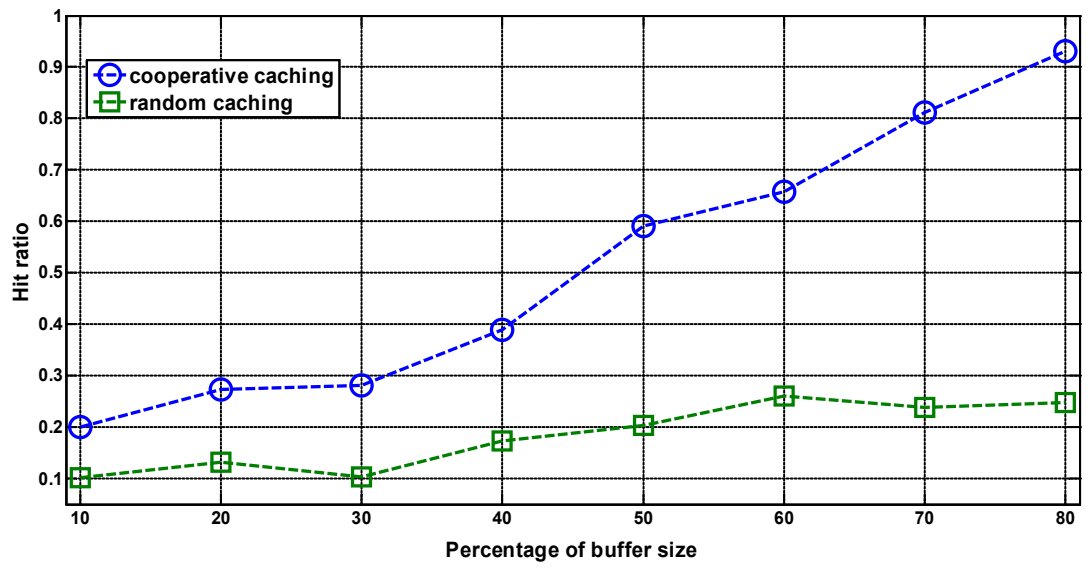

Fig. 5. Percentage of buffer size vs. hit ratio.

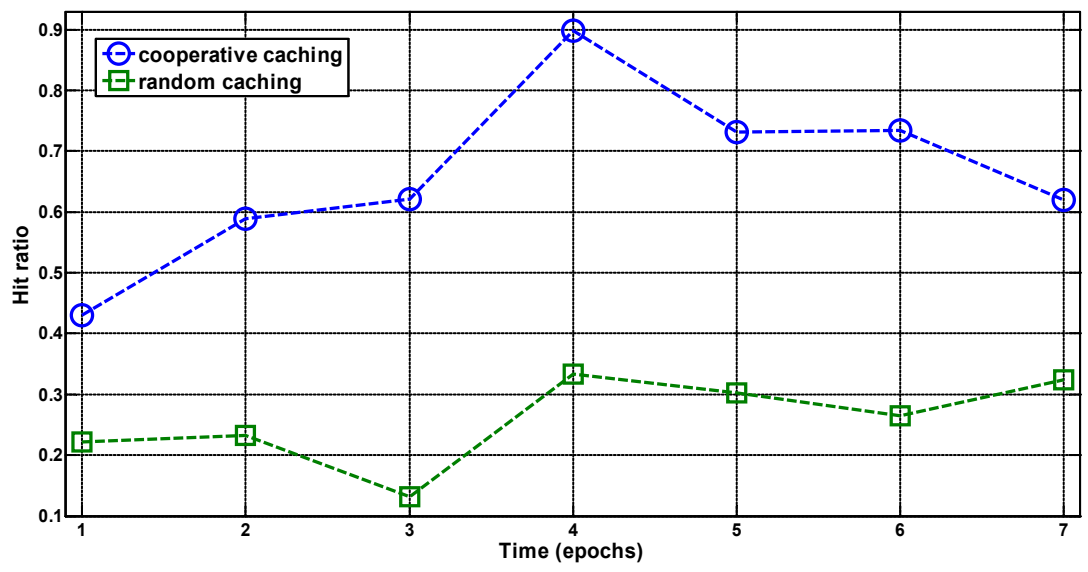

Fig. 6. Time vs. hit ratio.

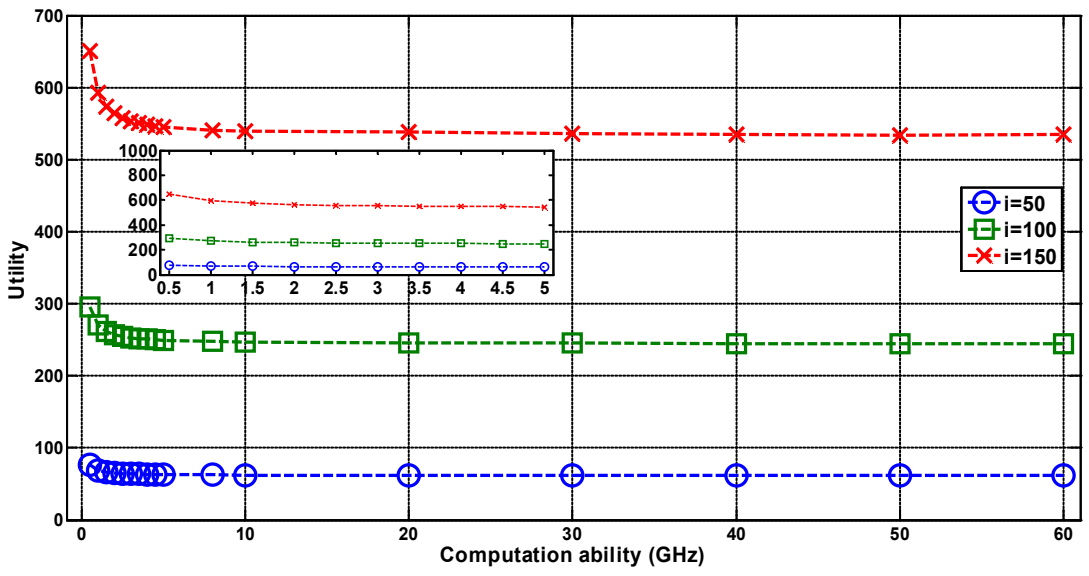

Fig. 7. Computation ability vs. utility.

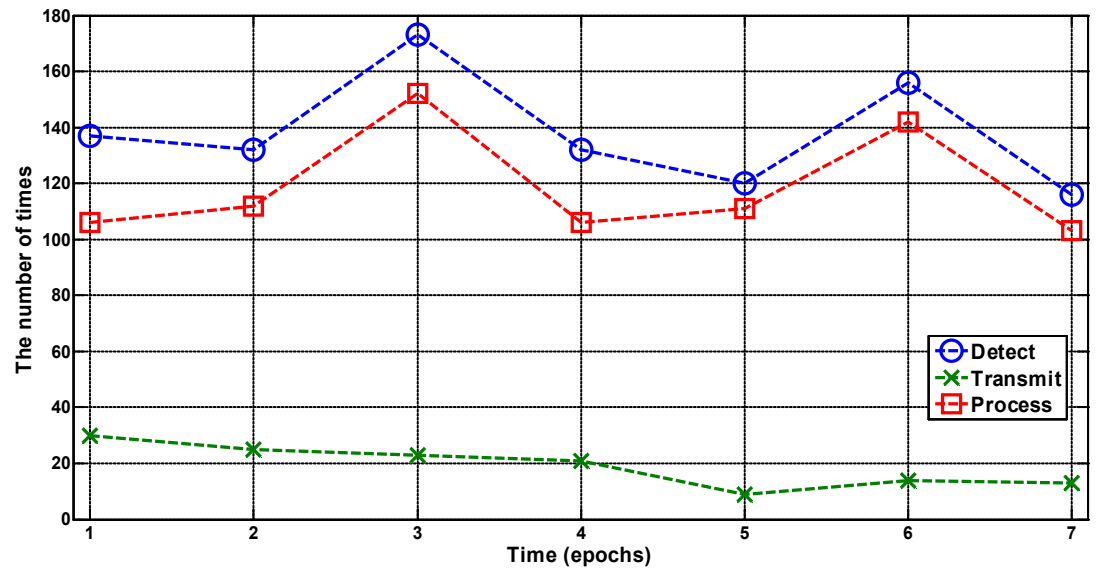

Fig. 8. Time vs. the number of times.

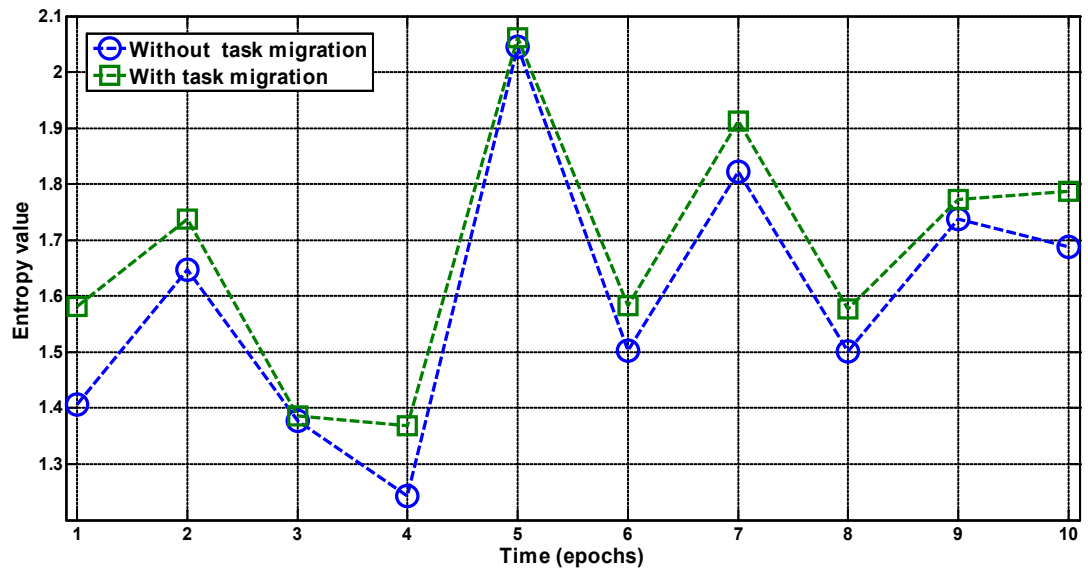

Fig. 9. Time vs. entropy value.

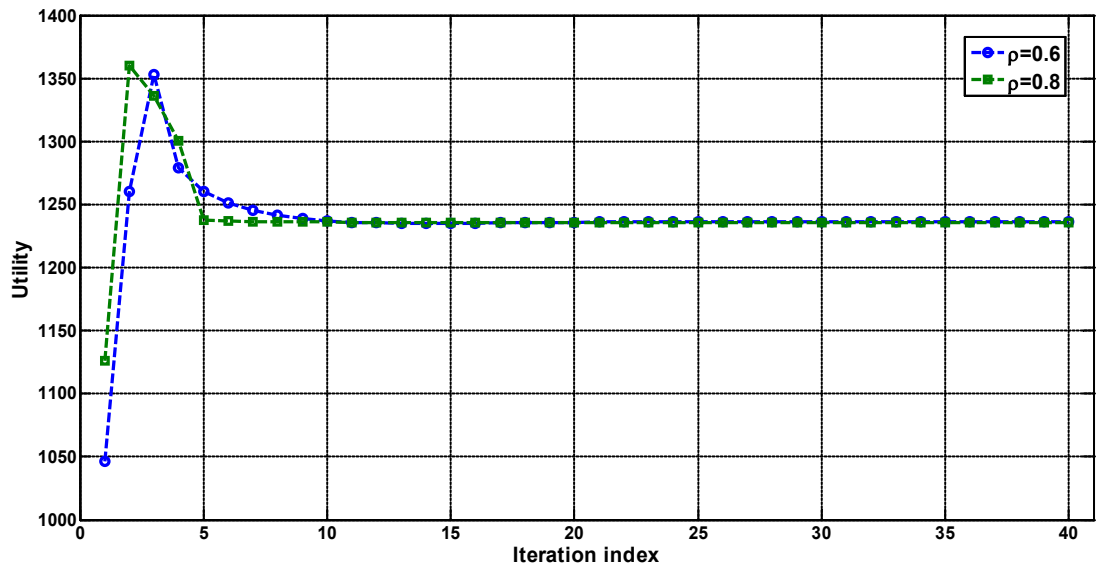

Fig. 10. Time vs. entropy value.

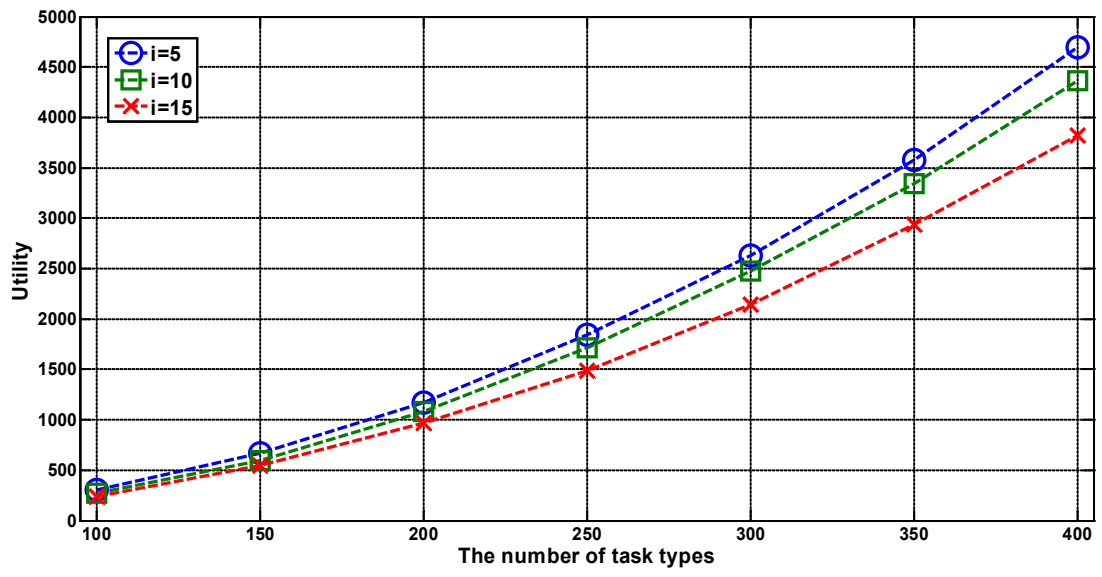

Fig. 11. The number of task types vs. utility.

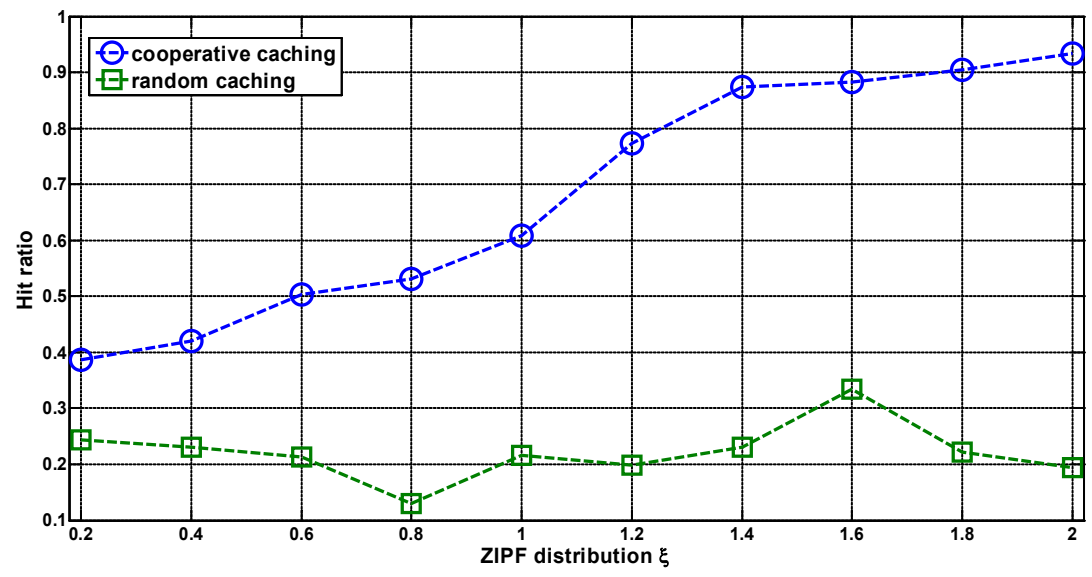

Fig. 12. The number of task types vs. utility.

The Fig.6 shows how the hit ration changed in cooperative caching and random caching respectively. The number of tasks is 5000, the type of tasks is 50, and the number of BS is 10. The hit ratio in cooperative caching is significantly more than that in random caching.

The Fig.7 represents the utility shows a downward trend with the computation ability increases from 0 GHz to 10 GHz. The utility shows a constant trend from 10 GHz to 60GHz. With increasing performance of computation ability and the utility will decrease gradually, but when the computation ability is more than 10GHz, the utility will remain stable. It is because that the number of tasks is limited and the higher performance of computation ability is not useful.

The Fig.8 shows the number of times in detecting, transmitting and processing respectively. The number of tasks is 5000, the type of tasks is 50, the number of BS is 10, and the number of groups is 5. When the group is in congestion, the transmitting action will be taken and transmit the tasks to the neighboring group, which is not in congestion and has the highest feedback. In each group, the proportion of transmitting is higher, which indicates that each group has the ability to perform task scheduling dynamically. With increase number of detecting tasks, the number of processing tasks will increase, and the number of transmitting tasks will also increase gradually.

The Fig.9 shows how the value of entropy changed with task migration and without task migration respectively. The number of tasks is 5000, the type of tasks is 50, and the number of BS is 10. The entropy value with task migration is slightly more than that without task migration. The reason is that each group has the ability to perform the task scheduling dynamically according to current environment.

In Fig.10, the number of tasks is 200, the type of tasks is 3, the number of BS is 3, and the computational capacity is 5GHz cycle/s. The rate of convergence in step 0.6 is slower than that in step 0.8. The step in 0.6 and 0.8 eventually converge in similar value. We can conclude that if the objective function is linear, we can get the optimal value. We verify the convergence of CL-ADMM algorithm in different step value.

The Fig.11 shows the increase trend in the utility in the different number of BS. The computational capacity is 5GHz cycle/s, and the step size is 1.The more the BS are, the lower the utility is.

The Fig.12 shows how the hit ratio changed in cooperative caching and random caching respectively. The number of tasks is 50000, the type of tasks is 50, and the number of BS is 10. The hit ratio in cooperative caching is significantly more than that in random caching. The hit ratio has a rise trend when the Zipf parameter changed between 0.2 and 2. It is because that with the increasing value of Zipf parameter, the type of tasks will be focused. It is means that the hit ratio will increase.

## VIII. CONCLUSION

In this paper, we present CL-ADMM framework for MEC resource management, which can leverage congestion in resource allocation among different groups, featuring distributed optimization and efficient caching mechanism, to reach intelligent and efficient resource management in MEC. It includes three parts: *cooperative task caching* based on semi-Markov process model, which aims at caching task efficiently; *dynamic task migration* based on cooperative improved Q-learning, which aims at effectively reducing delay and alleviating congestion; *task allocation* based on modified ADMM with Gaussian back substitution, which aims at minimizing delay and energy consumption for resources allocation in a group. The numerical results show that the proposed CL-ADMM framework can effectively reduce delay and energy consumption in MEC system.

## APPENDIX A
## PROOF OF THEOREM 1

*Proof:* The function $\psi(x,\varepsilon,\beta,\gamma,\delta)$ is strictly concave, so there exists an optimal solution. Suppose that $(\tilde{x},\tilde{\varepsilon},\tilde{\beta},\tilde{\gamma},\tilde{\delta})$ is an equilibrium point of the primal-dual algorithm for problem. Consider the candidate Lyapunov function (43) and suppose that $Q(\tilde{x},\tilde{\varepsilon},\tilde{\beta},\tilde{\gamma},\tilde{\delta})=0$, whenever $(x,\varepsilon,\beta,\gamma,\delta)\neq(\tilde{x},\tilde{\varepsilon},\tilde{\beta},\tilde{\gamma},\tilde{\delta})$, hence, we have (44).

$$Q(x,\varepsilon,\beta,\gamma,\delta)=\int_{\tilde{x}_{ij}}^{x_{ij}} w-\tilde{x}_{ij}\,dw+\int_{\tilde{\varepsilon}}^{\varepsilon}\theta-\tilde{\varepsilon}\,d\theta+\int_{\tilde{\beta}_{ij}}^{\beta_{ij}} v-\tilde{\beta}_{ij}\,dv+\int_{\tilde{\gamma}_{ij}}^{\gamma_{ij}}\tau-\tilde{\gamma}_{ij}\,d\tau+\int_{\tilde{\delta}_i}^{\delta_i}\eta-\tilde{\delta}_i\,d\eta \tag{43}$$

$$Q(x,\varepsilon,\beta,\gamma,\delta)=\frac{1}{2}(x_{ij}-\tilde{x}_{ij})^2+\frac{1}{2}(\varepsilon-\tilde{\varepsilon})^2+\frac{1}{2}(\beta_{ij}-\tilde{\beta}_{ij})^2+\frac{1}{2}(\gamma_{ij}-\tilde{\gamma}_{ij})^2+\frac{1}{2}(\delta_i-\tilde{\delta}_i)^2>0 \tag{44}$$

$$\dot{Q}(x,\varepsilon,\beta,\gamma,\delta)=(x_{ij}-\tilde{x}_{ij})\dot{x}_{ij}+(\varepsilon-\tilde{\varepsilon})\dot{\varepsilon}+(\beta_{ij}-\tilde{\beta}_{ij})\dot{\beta}_{ij}+(\gamma_{ij}-\tilde{\gamma}_{ij})\dot{\gamma}_{ij}+(\delta_i-\tilde{\delta}_i)\dot{\delta}_i \tag{45}$$

$$\begin{cases}\dot{x}_{ij}=\dfrac{d(-L_\rho(x,y^k,\lambda^k,v^k))}{dx}+\beta_{ij}-\gamma_{ij}-\varepsilon\left(\dfrac{u_j}{v(h_j)}+T_{execute}^{(b_i,h_j)}+T_{down}^{(b_i,h_j)}\right)-\delta_i c_j\\ \dot{\tilde{x}}_{ij}=\dfrac{d(-L_\rho(\tilde{x},y^k,\lambda^k,v^k))}{dx}+\beta_{ij}-\gamma_{ij}-\varepsilon\left(\dfrac{u_j}{v(h_j)}+T_{execute}^{(b_i,h_j)}+T_{down}^{(b_i,h_j)}\right)-\delta_i c_j=0\end{cases} \tag{46}$$

$$Q(y,\mu,\varsigma,\vartheta)=\int_{\tilde{y}_{ij}}^{x_{ij}}\pi-\tilde{y}_{ij}\,d\pi+\int_{\tilde{\mu}_{ij}}^{\mu_{ij}} o-\tilde{\mu}_{ij}\,do+\int_{\tilde{\varsigma}_{ij}}^{\varsigma_{ij}}\varpi-\tilde{\varsigma}_{ij}\,d\varpi+\int_{\tilde{\vartheta}_i}^{\vartheta_i}\kappa-\tilde{\vartheta}_i\,d\kappa \tag{47}$$

$$Q(y,\mu,\varsigma,\vartheta)=\frac{1}{2}(y_{ij}-\tilde{y}_{ij})^2+\frac{1}{2}(\mu_{ij}-\tilde{\mu}_{ij})^2+\frac{1}{2}(\varsigma_{ij}-\tilde{\varsigma}_{ij})^2+\frac{1}{2}(\vartheta_i-\tilde{\vartheta}_i)^2>0 \tag{48}$$

$$\dot{Q}(y,\mu,\varsigma,\vartheta)=(y_{ij}-\tilde{y}_{ij})\dot{y}_{ij}+(\mu_{ij}-\tilde{\mu}_{ij})\dot{\mu}_{ij}+(\varsigma_{ij}-\tilde{\varsigma}_{ij})\dot{\varsigma}_{ij}+(\vartheta_i-\tilde{\vartheta}_i)\dot{\vartheta}_i \tag{49}$$

Thus, we have $Q(x,\varepsilon,\beta,\gamma,\delta) > 0$ and (45).

According the KKT conditions, we can get

$$\begin{cases} \varepsilon \dot{\varepsilon} \le 0, \tilde{\varepsilon} \dot{\varepsilon} = 0 \\ \beta_{ij} \dot{\beta}_{ij} \le 0, \tilde{\beta}_{ij} \dot{\beta}_{ij} = 0 \\ \gamma_{ij} \dot{\gamma}_{ij} \le 0, \tilde{\gamma}_{ij} \dot{\gamma}_{ij} = 0 \\ \delta_i \dot{\delta}_i \le 0, \tilde{\delta}_i \dot{\delta}_i = 0 \end{cases}$$

Thus, we have (46) and

$$\begin{aligned} &\dot{Q}(x,\varepsilon,\beta,\gamma,\delta) \le (x_{ij} - \tilde{x}_{ij}) \dot{x}_{ij} \\ &= \left( \dot{x}_{ij} - \dot{\tilde{x}}_{ij} \right) (x_{ij} - \tilde{x}_{ij}) \\ &= 2\sigma(\tilde{x}_{ij} - x_{ij})(x_{ij} - \tilde{x}_{ij}) \le 0 \end{aligned}$$

Similarly, suppose that $(\tilde{y},\tilde{\mu},\tilde{\varsigma},\tilde{\vartheta})$ is the optimal solution. The Lyapunov function is (47), and we suppose that $Q(\tilde{y},\tilde{\mu},\tilde{\varsigma},\tilde{\vartheta})=0$ , whenever $(y,\mu,\varsigma,\vartheta) \ne (\tilde{y},\tilde{\mu},\tilde{\varsigma},\tilde{\vartheta})$ , hence, we have (48).

Thus, we have $Q(y,\mu,\varsigma,\vartheta) > 0$ and (49).

According the KKT conditions, we have

$$\begin{cases} \mu_{ij} \dot{\mu}_{ij} \le 0, \tilde{\mu}_{ij} \dot{\mu}_{ij} = 0 \\ \varsigma_{ij} \dot{\varsigma}_{ij} \le 0, \tilde{\varsigma}_{ij} \dot{\varsigma}_{ij} = 0 \\ \vartheta_i \dot{\vartheta}_i \le 0, \tilde{\vartheta}_i \dot{\vartheta}_i = 0 \end{cases} \tag{50}$$

Thus, we have (51) and (52).

$$\begin{cases} \dot{y}_{ij} = \dfrac{d(-L_\rho(x^{k+1},y^k,\lambda^k,z^k))}{dy^k} + \mu_{ij} - \varsigma_{ij} - \vartheta_i u_j \\ \dot{\tilde{y}}_{ij} = \dfrac{d(-L_\rho(x^{k+1},\tilde{y}^k,\lambda^k,z^k))}{dy} + \mu_{ij} - \varsigma_{ij} - \vartheta_i u_j = 0 \end{cases} \tag{51}$$

$$\begin{aligned} &\dot{Q}(y,\mu,\varsigma,\vartheta) \le (y_{ij} - \tilde{y}_{ij}) \dot{y}_{ij} \\ &= \left( \dot{y}_{ij} - \dot{\tilde{y}}_{ij} \right) (y_{ij} - \tilde{y}_{ij}) \\ &= 2\sigma(\tilde{y}_{ij} - \mathrm{y}_{ij})(y_{ij} - \tilde{y}_{ij}) \le 0 \end{aligned} \tag{52}$$

Thus, we can conclude that the proposed algorithm is following by the theory of Lyapunov, according to [28], and let $\{V_x^k\}$ and $\{V_y^k\}$ be the sequences generated by the proposed ADMM with Gaussian back substitution.

$$G = MO^{-1}M^T \tag{53}$$

Thus, we can obtain

$$\begin{aligned} \sum_{k=0}^{\infty} c_0 \left\| V_x^k - \tilde{V}_x^k \right\| \le \left\| V_x^0 - V_x^* \right\| \\ \sum_{k=0}^{\infty} c_0 \left\| V_y^k - \tilde{V}_y^k \right\| \le \left\| V_y^0 - V_y^* \right\| \end{aligned} \tag{54}$$

And thus we get

$$\begin{cases} \lim\limits_{k\to\infty} \left\| x_{ij}^k - \tilde{x}_{ij}^k \right\| = 0, \lim\limits_{k\to\infty} \left\| v_j^{k+1} - v_j^k \right\| = 0 \\ \lim\limits_{k\to\infty} \left\| y_{ij}^k - \tilde{y}_{ij}^k \right\| = 0, \lim\limits_{k\to\infty} \left\| z_j^{k+1} - z_j^k \right\| = 0 \end{cases} \tag{55}$$

According to the Variational Inequality, we have:

$$\begin{aligned} &\lim_{k\to\infty} (x_{ij} - \tilde{x}_{ij}^k) \left\{ \frac{dL_\rho(x,y^k,\lambda^k,v^k)}{dx} \right\} \ge 0 \\ &\lim_{k\to\infty} \left( \sum_{i=1}^{|BS|} x_{ij} - 1 \right) = 0 \quad j \in \{1,2,3...,|H|\} \\ &\lim_{k\to\infty} (y_{ij} - \tilde{y}_{ij}^k) \left\{ \frac{dL_\rho(x^{k+1},y^k,\lambda^k,v^k)}{dy} \right\} \ge 0 \\ &\lim_{k\to\infty} \left( \sum_{i=1}^{|BS|} y_{ij} - 1 \right) = 0 \;\; j \in \{1,2,3...,|H|\} \end{aligned} \tag{56}$$

## REFERENCES

[1] Y. Mao, C. You, J. Zhang, et. al., “A survey on mobile edge computing: The communication perspective,” *IEEE Communications Surveys and Tutorials* 19(4): 2322-2358, 2017.

[2] X. Sun, N. Ansari, “EdgeIoT: Mobile edge computing for the internet of things,” *IEEE Communications Magazine* 54(12): 22-29, 2016.

[3] N. Abbas, Y. Zhang, A. Taherkordi, T. Skeie, “Mobile edge computing: A survey,” *IEEE Internet of Things Journal* 5(1): 450-465, 2018.

[4] P. Zhao, G. Dán, “Time constrained service-aware migration of virtualized services for mobile edge computing,” *In proc. of IEEE ITC* (1) 2018: 64-72.

[5] S. Wang, R. Urgaonkar, M. Zafer, et.al., “Dynamic service migration in mobile edge computing based on markov decision process,” *IEEE/ACM Transactions on Networking* 27(3): 1272-1288, 2019.

[6] M. Chen, W. Li, G. Fortino, et. al, “A dynamic service migration mechanism in edge cognitive computing,” *ACM Transactions on Internet Technology* 19 (2): 30:1-30:15, 2019.

[7] C. Zhang, Z. Zheng, ”Task migration for mobile edge computing using deep reinforcement

learning," *Future Generation Computer Systems* 96: 111-118, 2019.

[8] L.Wang, L.Jiao, J. Li, et. al., "MOERA: Mobility-agnostic online resource allocation for edge computing," *IEEE Transactions on Mobile Computing*, 18(8): 1843-1856, 2019.

[9] M. Chen, Y. Hao, "Task offloading for mobile edge computing in software defined ultra-dense network," *IEEE Journal on Selected Areas in Communications* 36(3): 587-597, 2018.

[10] J. Xu, L. Chen, P. Zhou, "Joint service caching and task offloading for mobile edge computing in dense networks," *in Proc. of IEEE INFOCOM* 2018: 207-215.

[11] Z. Tan, F. Yu, X. Li, et. al., "Virtual resource allocation for heterogeneous services in full duplex-enabled SCNS with mobile edge computing and caching," *IEEE Transactions on Vehicular Technology* 67(2): 1794-1808, 2018.

[12] Y. Zhou, F. Yu, J. Chen, et al., "Resource allocation for information-centric virtualized heterogeneous networks with in-network caching and mobile edge computing," *IEEE Transactions on Vehicular Technology* 66(12): 11339-11351, 2017.

[13] Z. Zheng, L. Song, Z. Han, et al., A stackelberg game approach to proactive caching in large-scale mobile edge networks," *IEEE Transactions on Wireless Communications* 17(8): 5198-5211, 2018.

[14] Y. Wang, X. Tao, X. Zhang, et al., "Cooperative task offloading in three-tier mobile computing networks: An ADMM framework," *IEEE Transactions on Vehicular Technology* 68(3): 2763-2776, 2019.

[15] Z. Zhou, J. Feng, Z. Chang, et al., "Energy-efficient edge computing service provisioning for vehicular networks: A consensus ADMM approach," *IEEE Transactions on Vehicular Technology* 68(5): 5087-5099, 2019.

[16] Y. Dai, D. Xu, S. Maharjan, et al., "Joint computation offloading and user association in multi-task mobile edge computing," *IEEE Transactions on Vehicular Technology* 67(12): 12313-12325, 2018.

[17] P. Wang, Z. Zheng, B. Di, et al., "HetMEC: Latency-optimal task assignment and resource allocation for heterogeneous multi-layer mobile edge computing," *IEEE Transactions on Wireless Communications* 18(10): 4942-4956, 2019.

[18] C. Li, J. Tang, H. Tang, et al., "Collaborative cache allocation and task scheduling for data-intensive applications in edge computing environment," *Future Generation Computer Systems* 95: 249-264, 2019.

[19] H. Alameddine, S. Sharafeddine, S. Sebbah, et al., "Dynamic task offloading and scheduling for low-latency IoT services in multi-access edge computing," *IEEE Journal on Selected Areas in Communicati ons* 37(3): 668-682, 2019.

[20] C. Liu, M. Bennis, M. Debbah, et al., "dynamic task offloading and resource allocation for ultra-reliable low-latency edge computing," *IEEE Transactions on Communications* 67(6): 4132-4150, 2019.

[21] J. Meng, H. Tan, C. Xu, et al., "DEDAS: Online task dispatching and scheduling with bandwidth constraint in edge computing," in Proc. of *IEEE INFOCOM* 2019: 2287-2295.

[22] Y. Kim, J. Kwak, S. Chong, "Dual-side optimization for cost-delay tradeoff in mobile edge computing," *IEEE Transactions on Vehicular Technology* 67(2): 1765-1781, 2018.

[23] T. Ouyang, R. Li, X. Chen, et al., "Adaptive user-managed service placement for mobile edge computing: An online learning approach," *in proc. of IEEE INFOCOM* 2019: 1468-1476.

[24] L. Tan, Q. Hu, "Mobility-aware edge caching and computing in vehicle networks: A deep reinforcement learning," *IEEE Transactions on Vehicular Technology* 67(11): 10190-10203, 2018.

[25] Y. Sun, X. Guo, J. Song, et al., "Adaptive learning-based task offloading for vehicular edge computing systems," *IEEE Transactions on Vehicular Technology* 68(4): 3061-3074, 2019.

[26] W. Chien, H. Weng, C. Lai, "Q-learning based collaborative cache allocation in mobile edge computing," *Future Generation Computer Systems* 102: 603-610, 2020.

[27] M. Liu, F. Yu, Y. Teng, et. al., "Computation offloading and content caching in wireless blockchain networks with mobile edge computing," *IEEE Transactions on Vehicular Technology* 67(11): 11008-11021, 2018.

[28] B. He, M. Tao, X. Yuan, "Alternating direction method with Gaussian back substitution for separable convex programming," *SIAM J. Optim.* 22, 313-340, 2012.

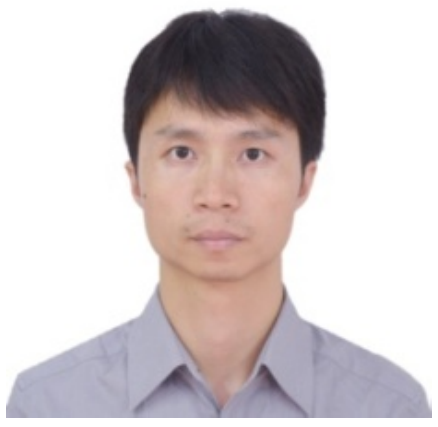

**Xiaoxiong Zhong** (S'12-M'16) received his Ph.D degree in Computer Science and Technology from Harbin Institute of Technology, China, in 2015. He was a Postdoctoral Research Fellow with Tsinghua University, from 2016 to 2018. He is currently an assistant professor with the Cyberspace Security Research Center, Peng Cheng Laboratory, Shenzhen, China. His general research interests include network protocol design and analysis, data transmission and data analysis in internet of things and edge computing.

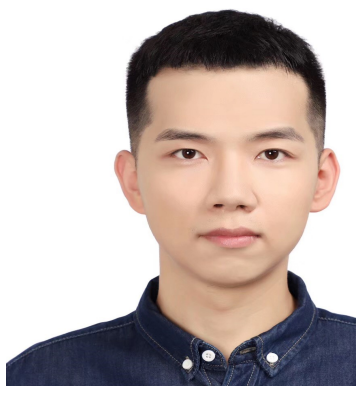

**Xinghan Wang** received his B.S. degree in Computer Science and Technology from Taiyuan University of Technology, China, in 2016. He is currently pursuing the M.S. degree in the School of Computer Science and Information Security at Guilin University of Electronic Technology, China. His research interest includes reinforcement learning and edge computing.

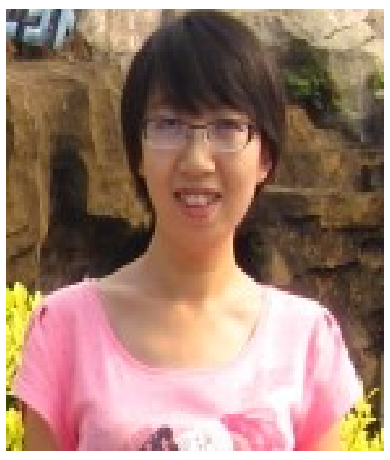

**Li Li** received her Ph.D degree in Computer Science and Technology from Harbin Institute of Technology, China, in 2017. She is currently a postdoctoral fellow with the Graduate School at Shenzhen, Tsinghua University, China. Her research interest is in the area of wireless networks and opportunistic networks.

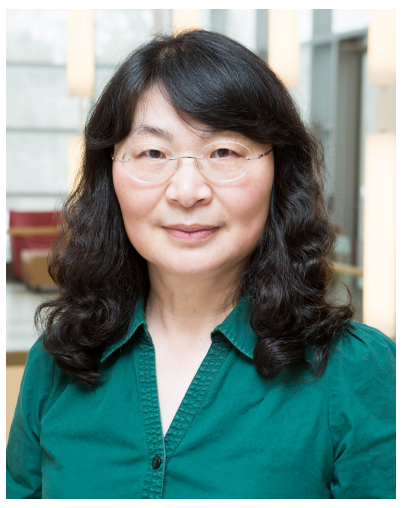

**Yuanyuan Yang** received the BEng and MS degrees in computer science and engineering from Tsinghua University and the MSE and PhD degrees in computer science from Johns Hopkins University. She is a SUNY Distinguished Professor of computer engineering and computer science and the Associate Dean for Academic Affairs in the College of Engineering and Applied Sciences at Stony Brook University, New York. Her research interests include wireless networks, data center networks and cloud computing. She has published over 380 papers in major journals and refereed conference proceedings and holds seven US patents in these areas. She is currently the Associate Editor-in-Chief for IEEE Transactions on Cloud Computing and an Associate Editor for ACM Computing Surveys. She has served as an Associate Editor-in-Chief and Associated Editor for IEEE Transactions on Computers and Associate Editor for IEEE Transactions on Parallel and Distributed Systems. She has also served as a general chair, program chair, or vice chair for several major conferences and a program committee member for numerous conferences. She is an IEEE Fellow.

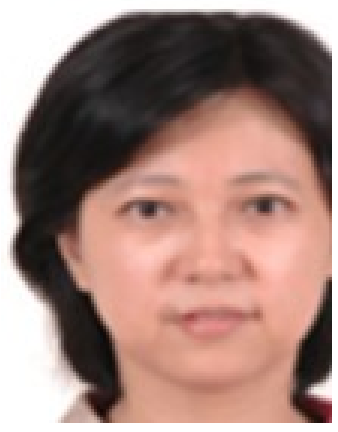

**Yang Qin** (S'98-M'01-SM'06) received her B.S (with first class honors) in Computer Science at Southwest Jiaotong University (China), in 1989, M.S in Computer Science at Huazhong University of Science & Technology, Wuhan, Hubei, in 1992, and Ph.D in Computer Science, Hong Kong University of Science & Technology, Kowloon, Hong Kong at November of 1999. From 1999 to 2000, she has visited the Washington State University as a Postdoc, USA. From 2000 to 2008, she is an assistant professor Nanyang Technological University, Singapore. Currently, she is an associate professor in the Department of Computer Science and Technology, Harbin Institute of Technology (Shenzhen), China. She is a senior member of IEEE.

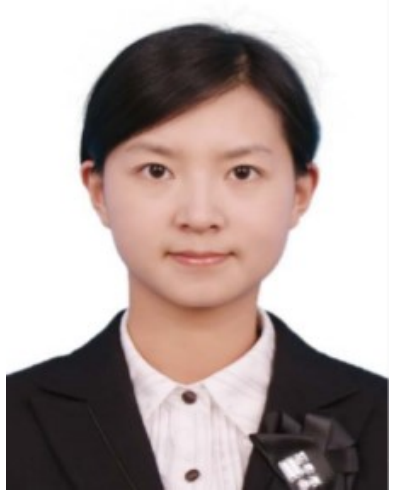

**Tingting Yang** (M'13) received the Ph.D. degrees from Dalian Maritime University, China, in 2010. She is currently a professor in the School of Electrical Engineering and Intelligentization, Dongguan University of Technology, China. Since September 2012, she has been a visiting scholar at the Broadband Communications Research (BBCR) Lab at the Department of Electrical and Computer Engineering, University of Waterloo, Canada. Her research interests are in the areas of maritime wideband communication networks, DTN networks, and green wireless communication.

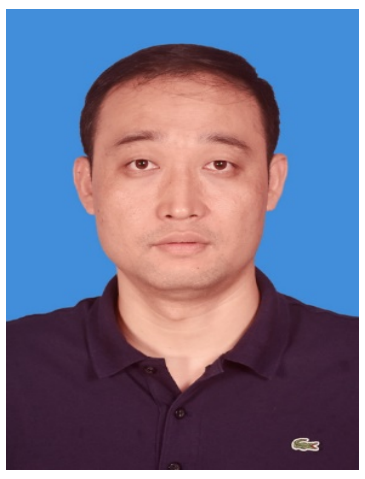

**Bin Zhang** received his Ph.D. degree in Department of Computer Science and Technology, Tsinghua University, China in 2012. He worked as a post doctor in Nanjing Telecommunication Technology Institute from 2014 to 2017. He is now a researcher in Peng Cheng Laboratory, Shenzhen, China. His current research interests focus on network anomaly detection information privacy security, etc.

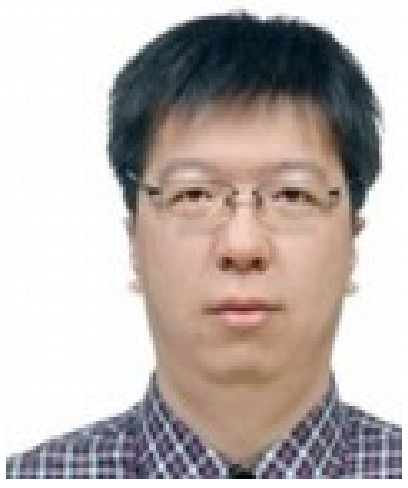

**Weizhe Zhang** (SM'15) is currently a professor in the School of Computer Science and Technology at Harbin Institute of Technology, China, and the director in the Cyberspace Security Research Center, Peng Cheng Laboratory, Shenzhen, China. His research interests are primarily in cyberspace security, cloud computing, and high-performance computing. He has published more than 130 academic papers in journals, books, and conference proceedings. He is a senior member of the IEEE and a lifetime member of the ACM.